\documentclass[preprint]{aastex}

\usepackage{graphicx,amssymb,amsmath,multirow,gensymb,lscape}
\usepackage{enumitem}
\usepackage{natbib}
\newcommand{\vol}{\mbox{cm$^{-3}$}} 
\newcommand{\cden}{\mbox{cm$^{-2}$}} 
\newcommand{\kms}{\mbox{km s$^{-1}$}}
\newcommand{\um}{\mbox{$\mu$m}}
\newcommand{\ammonia}{\mbox{NH$_{3}$}}
\newcommand{\Msun}{\mbox{M$_{\odot}$}}
\newcommand{\cmg}{\mbox{cm$^2$ g$^{-1}$}}
\newcommand{\Av}{\mbox{$A_V$}}
\newcommand{\NHH}{\mbox{N(H$_2$)}}
\newcommand{\nHH}{\mbox{n(H$_2$)}}
\newcommand{\NCCO}{\mbox{N(\mbox{$^{13}$CO})}}
\newcommand{\NCOO}{\mbox{N(\mbox{C$^{18}$O})}}
\newcommand{\CO}{\mbox{$^{12}$CO}}
\newcommand{\CCO}{\mbox{$^{13}$CO}}
\newcommand{\COO}{\mbox{C$^{18}$O}}

\newcommand{\XCOO}{\mbox{X(C$^{18}$O)}}

\newcommand{\Tex}{\mbox{$T_{ex}$}}
\newcommand{\Tmb}{\mbox{$T_{MB}$}}

\newcommand{\kmspc}{\mbox{km s$^{-1}$ pc$^{-1}$}}
\newcommand{\Kcm}{\mbox{K cm$^{-3}$}}

\shorttitle{Kinematic and Chemical Properties of B1-E}
\shortauthors{Sadavoy et al.}

\title{The Kinematic and Chemical Properties of a Potential Core-Forming Clump:  Perseus B1-E}
\author{S. I. Sadavoy\altaffilmark{1}, Y. Shirley\altaffilmark{2,1}, J. Di Francesco\altaffilmark{3}, Th. Henning\altaffilmark{1}, M. J. Currie\altaffilmark{4}, Ph. Andr\'{e}\altaffilmark{5}, S. Pezzuto\altaffilmark{6}}
\altaffiltext{1}{Max-Planck-Institut f\"{u}r Astronomie (MPIA), K\"{o}nigstuhl 17, D-69117 Heidelberg, Germany}
\altaffiltext{2}{Astronomy Department, The University of Arizona, 933 N. Cherry Ave., Tucson, AZ 85721, USA}
\altaffiltext{3}{National Research Council Canada, 5071 West Saanich Road, Victoria BC Canada, V9E 2E7}
\altaffiltext{4}{Joint Astronomy Centre, 660 N. A$^{\prime}$oh\={o}k\={u} Place, University Park, Hilo, Hawaii, 96720, USA}
\altaffiltext{5}{Laboratoire AIM, CEA/DSM-CNRS-Universit\'{e} Paris Diderot, IRFU/Service dÕAstrophysique, Saclay, 91191 Gif-sur-Yvette, France}
\altaffiltext{6}{Istituto di Astrofisica e Planetologia Spaziali, via Fosso del Cavaliere 100, 00133, Rome, Italy}

\begin{document}

\begin{abstract}

We present \CCO\ and \COO\ (1-0), (2-1), and (3-2) maps towards the core-forming Perseus B1-E clump using observations from the James Clerk Maxwell Telescope (JCMT), Submillimeter Telescope (SMT) of the Arizona Radio Observatory, and IRAM 30 m telescope.  We find that the \CCO\ and \COO\ line emission both have very complex velocity structures, indicative of multiple velocity components within the ambient gas.  The (1-0) transitions reveal a radial velocity gradient across B1-E of $\sim 1\ \kmspc$ that increases from north-west to south-east, whereas the majority of the Perseus cloud has a radial velocity gradient increasing from south-west to north-east.  In contrast, we see no evidence of a velocity gradient associated with the denser \emph{Herschel}-identified substructures in B1-E.  Additionally, the denser substructures have much lower systemic motions than the ambient clump material, which indicates that they are likely decoupled from the large-scale gas.  Nevertheless, these substructures themselves have broad line widths ($\sim 0.4$ \kms) similar to that of the \COO\ gas in the clump, which suggests they inherited their kinematic properties from the larger-scale, moderately dense gas.  Finally, we find evidence of \COO\ depletion only toward one substructure, B1-E2, which is also the only object with narrow (transonic) line widths.  We suggest that as prestellar cores form, their chemical and kinematic properties are linked in evolution, such that these objects  must first dissipate their turbulence before they deplete in CO.

\end{abstract}


\section{Introduction\label{Intro}}

Stars form within prestellar cores, which are small ($\lesssim 0.1$ pc), cold ($\lesssim 10$ K), and dense ($\gtrsim 10^{5}$ \vol) objects inside molecular clouds (e.g., \citealt{Williams00}; \citealt{difran07}).  In addition, prestellar cores are dynamically quiescent both in their internal velocity dispersions (e.g., \citealt{Rosolowsky08}; \citealt{Pineda10}) and in their systemic motions (e.g., \citealt{Kirk10}), and most prestellar cores have similar centroid velocities between their densest, central material and their moderately dense envelope (e.g., \citealt{Walsh04}; \citealt{Kirk07}; \citealt{Andre07}).  These observations suggest that (1) prestellar cores have homogenized their relative motions compared to the surrounding, ambient cloud and (2) the densest gas material within these cores are coupled to (e.g., move quiescently with) their surrounding envelope.  

Prestellar cores also have significant chemical differentiation, where the higher density environments are best probed with nitrogen-bearing molecules or deuterated tri-hydrogen molecules and the lower density envelopes are best traced with carbon-bearing molecules (e.g., see \citealt{difran07}).   This differentiation can be attributed to different effective excitation densities between molecules (\citealt{Shirley15}) and to different temperatures for molecular ``freeze-out'', where certain molecular species adsorb onto dust grains and deplete from the gas phase.  In particular, carbon-bearing molecules adsorb easily in cold, dense prestellar cores (e.g., \citealt{Bergin95, Bergin02}; \citealt{Kramer99}; \citealt{Caselli99}), and observations of such objects often show absorption at 4.67 \um\ attributed to CO-ices (e.g., \citealt{Chiar94}).  Additionally, the formation of carbon-rich icy mantles on dust grains  provides a surface for larger, organic molecules to form via the irradiation of those icy mantles (e.g., see \citealt{Herbst09}).

Therefore, prestellar cores have distinct dynamical and chemical signatures compared with their parent cloud.  Most clouds are actively forming stars or contain a current generation of dense cores, and thus, can no longer be used to constrain the initial conditions for their formation.  As a result, simulations and models of molecular clouds often assume a wide range of physical conditions, such as radiation feedback (e.g., \citealt{Bate09}; \citealt{Offner09}; \citealt{Krumholz10}), magnetic fields (e.g., \citealt{LiNakamura04}; \citealt{Basu09,Basu09b}), different turbulent properties (e.g., \citealt{Klessen00}; \citealt{Bate09b}; \citealt{Heitsch11}), or a combination of these effects (e.g., \citealt{PriceBate09}; \citealt{Commercon11}; \citealt{Padoan11}).  Thus, observations of a young core-forming region are key to constraining better the physical processes that cause  cores to condense out of the larger-scale ambient cloud material.

The B1-E clump in the Perseus molecular cloud is an ideal laboratory to study core formation in relative isolation.   B1-E is the sole dense ($\Av > 5$ mag) region in Perseus without young stellar objects (YSOs; e.g., \citealt{Evans09}) and prestellar cores (e.g., \citealt{Enoch06}; \citealt{difran08}).  Nevertheless, using \emph{Herschel} data, \citet{Sadavoy12} identified nine distinct substructures in B1-E.  Follow-up \ammonia\ (1,1) observations of these substructures showed that most of them had supersonic non-thermal motions with only one source, B1-E2, having more quiescent, transonic non-thermal motions.  Since the \ammonia\ emission from these substructures was weaker than the  typical line peaks associated with prestellar cores (e.g., \citealt{Rosolowsky08}), the B1-E substructures appear to be in an unique state where their dense gas is both minimal and turbulent.   Thus, the B1-E clump is an excellent candidate for a core-forming region that is fragmenting for the first time, such that its \emph{Herschel}-identified substructures represent early-stage core precursors. 

For this paper, we mapped the entire B1-E clump in \CCO\ and \COO\  emission of their $J=1-0$, $J=2-1$, and $J=3-2$ transitions.  We use these data to characterize the kinematic and chemical properties of the B1-E clump and its substructures.   This paper is outlined as follows: in Section \ref{data}, we present our \CCO\ and \COO\ observations; in Section \ref{results}, we describe the kinematic properties of B1-E and its substructures using our line observations; in Section \ref{radex}, we use radiative transfer modeling to determine the chemical properties across B1-E; in Section \ref{discussion}, we discuss the observed kinematic and chemical signatures, and we propose an evolutionary picture for core formation; and finally, in Section \ref{conc}, we present our conclusions and summarize the paper.

\section{Data}\label{data}

\subsection{Herschel}

The Perseus molecular cloud was observed by \emph{Herschel} as part of the \emph{Herschel} Gould Belt Survey\footnote{http://gouldbelt-herschel.cea.fr/} (HGBS; \citealt{Andre10}), with the Western field mapped in February 2010 and the Eastern field mapped in February 2011.  Both Perseus fields were observed using the PACS/SPIRE parallel mode scanning technique with a 60 arcsec s$^{-1}$ scan rate (\citealt{Sadavoy12}; \citealt{Pezzuto12}; Pezzuto et al. in preparation).  The data were reduced using HIPE version 7 and modified standard reduction scripts by M. Sauvage (PACS) and P. Panuzzo (SPIRE) in the same manner as \citet{Sadavoy14}.  We used \emph{scanamorphos} 11 (\citealt{Roussel12}) to produce the final maps.    

We corrected the zero-point offset at each wavelength for both the Eastern and Western fields using Planck observations following  \citet{Bernard10}.  Since both fields maps overlapped toward B1-E, we mosaicked the two fields using the \textsc{Starlink} task \texttt{wcsmosaic} and bilinear interpoloation.  For the $160 - 500\ \um$\ data, the two fields had excellent agreement in their mosaics.   Finally, we convolved the mosaics to a common resolution of 36.3\arcsec\ (the 500 \um\ beam) and a common grid of 14\arcsec\ pixels.

\subsection{James Clerk Maxwell Telescope}

We observed the B1-E clump with the Heterodyne Array Receiver Programme (HARP) detector at the James Clerk Maxwell Telescope (JCMT) in the same manner as the JCMT Gould Belt Survey (\citealt{Buckle09}).  The HARP detector was tuned to map B1-E simultaneously in \CCO\ (3-2) and \COO\ (3-2) line emission at 330.58796 GHz and 329.330545 GHz, respectively.  Observations were taken using position-switching, orthogonal raster scans with 1/4 array spacing over a $25\arcmin \times 14\arcmin$ region.  The backend was configured to 4096 channels over 250 MHz, resulting in $\sim 0.055$ \kms\ velocity resolution.  B1-E was observed on 13 July 2011 and 12 January 2012 for $\sim 2.2$ hours of observing time ($\sim 20$ \%\ of the total project time).  We selected the coordinates 03:35:47.7, 31:45:30.5 (J2000.0) for our off-position.

The HARP data were processed using the Auto-Correlation Spectral Imaging System (ACSIS; \citealt{Jenness08}) and reduced using the ACSIS pipeline (\citealt{Cavanagh08}) in a similar manner as described in \citet{Currie13} and \citet{Sadavoy13}.  In brief, the HARP data were reduced using a two-stage, fully automated methodology.  First, we conducted a quality assurance, which mostly removed those spectra degraded by a variety of instrumental effects.  Second, we performed an iterative procedure that used spectral cubes to estimate linear baselines and to identify emission features to be masked for a more refined baseline subtraction.  For the B1-E data, one iteration was sufficient to characterize the baselines.  Detailed descriptions of the methods used can be found in Jenness et al. (2015, submitted).

Finally, we used a main-beam efficiency of $\eta_{MB} = 0.61$ for our reduced spectra.  The effective beam of these data is 17.3\arcsec\ , which we convolved with a 31.9\arcsec\ beam to get our data at an equivalent resolution of 36.3\arcsec, or the resolution of our 500 \um\ \emph{Herschel} observations.  The 1 $\sigma$ line sensitivities at this spatial resolution are $\sim 0.2$ K and $\sim 0.3$ K for the \CCO\ (3-2) and \COO\ (3-2) data, respectively.

\subsection{Submillimeter Telescope}\label{smtObs}

The B1-E clump was mapped using the 10 m of the Arizona Radio Observatory (ARO) Submillimeter Telescope (SMT) in dual polarization, 4-IF mode over $\sim 35$ hours in March 2012. We used the 1 mm ALMA prototype receivers to make position-switching on-the-fly (OTF) maps with the upper sideband tuned to \CO\ (2-1) at 230.538 GHz and the lower sideband tuned to \CCO\ (2-1) at 220.3987 GHz with a $\sim$ 5.0 GHz IF. With the 250 kHz filterbank backends, the velocity resolution is $\sim 0.34$ \kms\ and the spatial resolution is $\sim 33.7$\arcsec\ at 220 GHz.  At this time, we only discuss the \CCO\ (2-1) data.

Point and focus observations were made on Venus or Jupiter.  Line calibrations to measure the sideband spillover were made using the standard SMT line calibrator in Orion A (J2000.0: 05:35:14.48, -05:22:27.6).  The beam efficiency ($\eta \approx 0.62$) was measured using the average efficiency from observations of Venus and Jupiter.  The beam efficiency was stable (within 5\%) throughout the observations.

To avoid artifacts introduced by long scans, we divided B1-E into smaller subregions, mapping each block separately. We used four overlapping subregions, where each subregion was $9\arcmin\ \times 7\arcmin$, with a 2\arcmin\ overlap, for a total map area of $\sim 190$ arcmin$^2$.  We initially used an OTF scanning rate of 10\arcsec/s, but switched to 20\arcsec/s to speed up the mapping. Each region was observed in orthogonal scans (along RA or decl.) with 10\arcsec\ spacings at least once at 10\arcsec/s or twice at 20\arcsec/s along each direction.  We selected the coordinates 03:35:31.0, 31:45:31.0 (J2000.0) for our off-position. 

The observations were reduced using the \textsc{gildas} reduction software, \textsc{class}.  We reduced the horizontal and vertical polarizations separately using an initial single-order baseline fit.  The telescope was fairly stable over the observations, and we rejected only two scans in one subregion due to unusual spectral signatures.  The spectral line data were combined onto a evenly spaced grid and smoothed to 36.3$\arcsec$ resolution, the resolution at the 500 \um\ beam.  Several rows of spectra had curved baselines.  To improve these baselines, we subtracted an additional quadratic function using a wide velocity range (-11 \kms\ to 28 \kms).  The final 1 $\sigma$ rms of the smoothed maps is $\sim 0.05$ K towards the center (highest sampling) and $\sim 0.15$ K towards the corners of the map.

\subsection{IRAM 30 m Telescope}

We also fully mapped B1-E with the IRAM 30 m telescope in August 2012 using the EMIR receiver and VESPA backend to observe simultaneously  \CCO\ (1-0) at 110.2014 GHz, \COO\ (1-0) at 109.7822 GHz, and \COO\ (2-1) at 219.5604 GHz.  We made frequency-switching OTF maps with a throw of 3.9 MHz in the 90 GHz band and 11.7 MHz in the 230 GHz band.  The respective velocity resolutions are $\sim 0.106$ \kms\ and $\sim 0.053$ \kms\ and the respective spatial resolutions are $\sim 24$\arcsec\ and $\sim 12$\arcsec\ at 110 GHz and 220 GHz. 

Like our SMT observations, we mapped B1-E with orthogonal scans in four smaller, overlapping subregions for a total coverage of $\sim 190$ arcmin$^2$ in all three tracers (see Section \ref{smtObs}) with a step size between scans of 5\arcsec\ and a scan rate of 7.5\arcsec/sec.  Point and focus observations were done using Uranus after full scans of each subregion.  Line calibrators were observed while B1-E was in transit.  We adopted beam efficiencies and forward efficiencies of $\eta_{beam} = 0.78$ and $F_{eff} = 0.94$ at 110 GHz and $\eta_{beam} = 0.61$ and $F_{eff} = 0.93$ at 220 GHz.  The total on sky time was $\sim 13$ h.  

The IRAM 30 m observations were reduced using \textsc{class}.  We removed an initial single-order baseline from all spectra before folding the spectra.  For the \COO\ (1-0) and \COO\ (2-1) observations, we subtracted additional single-order or second-order baselines to remove large-scale ripples after the spectra were folded.  As with the previous line data, we smoothed the data to a similar grid as our the 500 \um\ \emph{Herschel} data.  The final 1 $\sigma$ rms of the smoothed maps range from $\sim 0.03 - 0.1$ K for \CCO\ (1-0), $\sim 0.03 - 0.1$ K for \COO\ (1-0), and $\sim 0.1 - 0.2$ K for \COO\ (2-1) between the center of the map (highest sampling) and the corners (lowest sampling).

\section{Results}\label{results}

\begin{figure}[h!]
\includegraphics[scale=0.75]{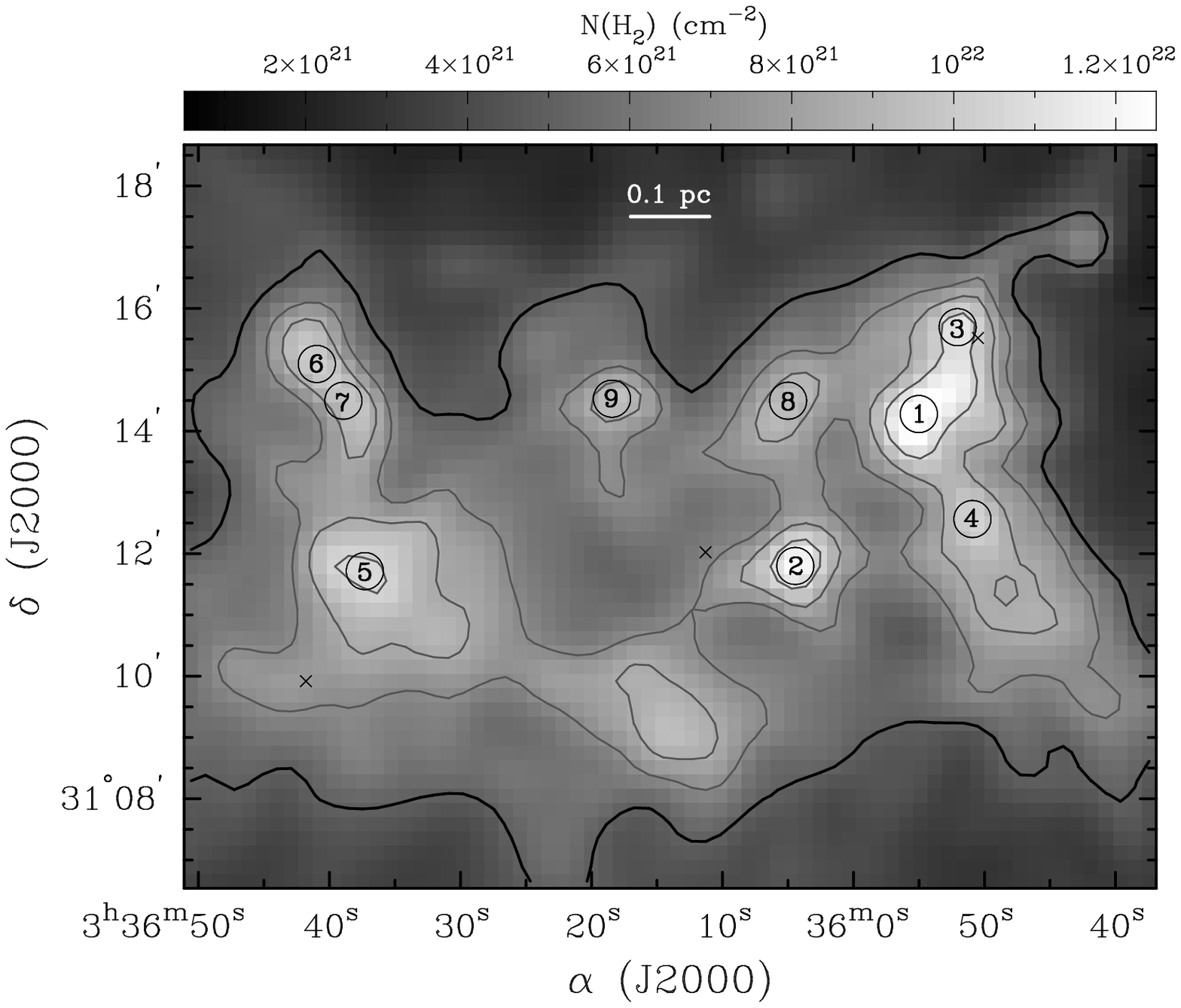}
\caption{\emph{Herschel}-derived column density map for B1-E based on modified blackbody fits to the $160-500$ \um\ SEDs.  The position of the nine substructures from \citet{Sadavoy12} are also given, with circles corresponding to the map resolution (36.3\arcsec).  The grey contours correspond to column density levels of $7.0 \times 10^{21}\ \cden, 8.5 \times 10^{21}\ \cden, \mbox{and}\ 10.5 \times 10^{21}$ \cden, and the thick black contour represents the outer boundary of the clump ($\NHH = 5.0 \times 10^{21}\ \cden$; $\Av \sim 7$ mag).  The small crosses correspond to spectra shown in Figure \ref{sampleSpec_13co}. \label{b1e_colDen}}
\end{figure}

Figure \ref{b1e_colDen} shows the \emph{Herschel}-derived column density map for the B1-E clump.  Column densities (and line-of-sight dust temperatures) were determined from modified blackbody fits to the \emph{Herschel} data in the same manner as \citet{Sadavoy14}, where we assumed a fixed dust opacity law corresponding to $\beta = 2$ and $\kappa_0 = 0.1$ \cmg\ at 300 \um, which is the adopted dust opacity law by the HGBS consortium (e.g., see \citealt{Roy14}), and we applied average colour corrections to the \emph{Herschel} field.  Moreover, we assumed a mean molecular weight per hydrogen molecule of $\mu = 2.8$ (\citealt{Kauffmann08}).  Note that \citet{Sadavoy12} adopted a mean molecular weight per unit mass ($\mu = 2.33$).  Additionally, the observations in \citet{Sadavoy12} were reduced using HIPE 5, and due to updated calibration references, the PACS fluxes changed by $\lesssim 20\%$\ between HIPE 5 and HIPE 6 (B. Altieri 2013, private communication).  Thus, the column densities presented here are lower by $\sim 32$\%\ compared to those from \citet{Sadavoy12}. 

Figure \ref{b1e_colDen} also gives the positions of the nine B1-E substructures identified by \citet[][see paper for coordinates]{Sadavoy12}.  These substructures are typically $< 0.1$ pc in size with masses of $\lesssim 1$ \Msun, and average densities of $\sim 10^4$ \vol.  For more physical information about the B1-E substructures, see \citet{Sadavoy12}.

Table \ref{b1eSum} lists the mass, approximate angular dimensions, area, effective radius, and median temperature of the B1-E clump using the \emph{Herschel} data alone.  These values correspond only to the region of B1-E with $\NHH > 5 \times 10^{21}$ \cden\ (thick black contour in Figure \ref{b1e_colDen}).   Mass was measured from the column densities, clump area was determined from the number of pixels above the column density threshold, and the effective radius was measured assuming B1-E is circular, $R_{eff} = \sqrt{A/\pi}$, assuming a distance of 235 pc (\citealt{Hirota08}).   The median temperature comes from the line-of-sight dust temperatures determined from the modified blackbody fits to the \emph{Herschel} data.  

\begin{table}[h!]
\caption{\emph{Herschel}-derived B1-E Properties}\label{b1eSum}
\begin{tabular}{ccccc}
\hline\hline
$M$ 	(\Msun) 	& Angular Size 	& $A$ (pc$^2$)	&	$R_{eff}$	(pc)&	$T$ (K)\\
\hline
88			& $15.5\arcmin \times 10.5\arcmin$		& 0.57		&	0.43		&	13.8 \\
\hline
\end{tabular}
\end{table}

\subsection{Velocity Structure}\label{vel_structure}

\begin{figure}[h!]
\includegraphics[scale=0.75]{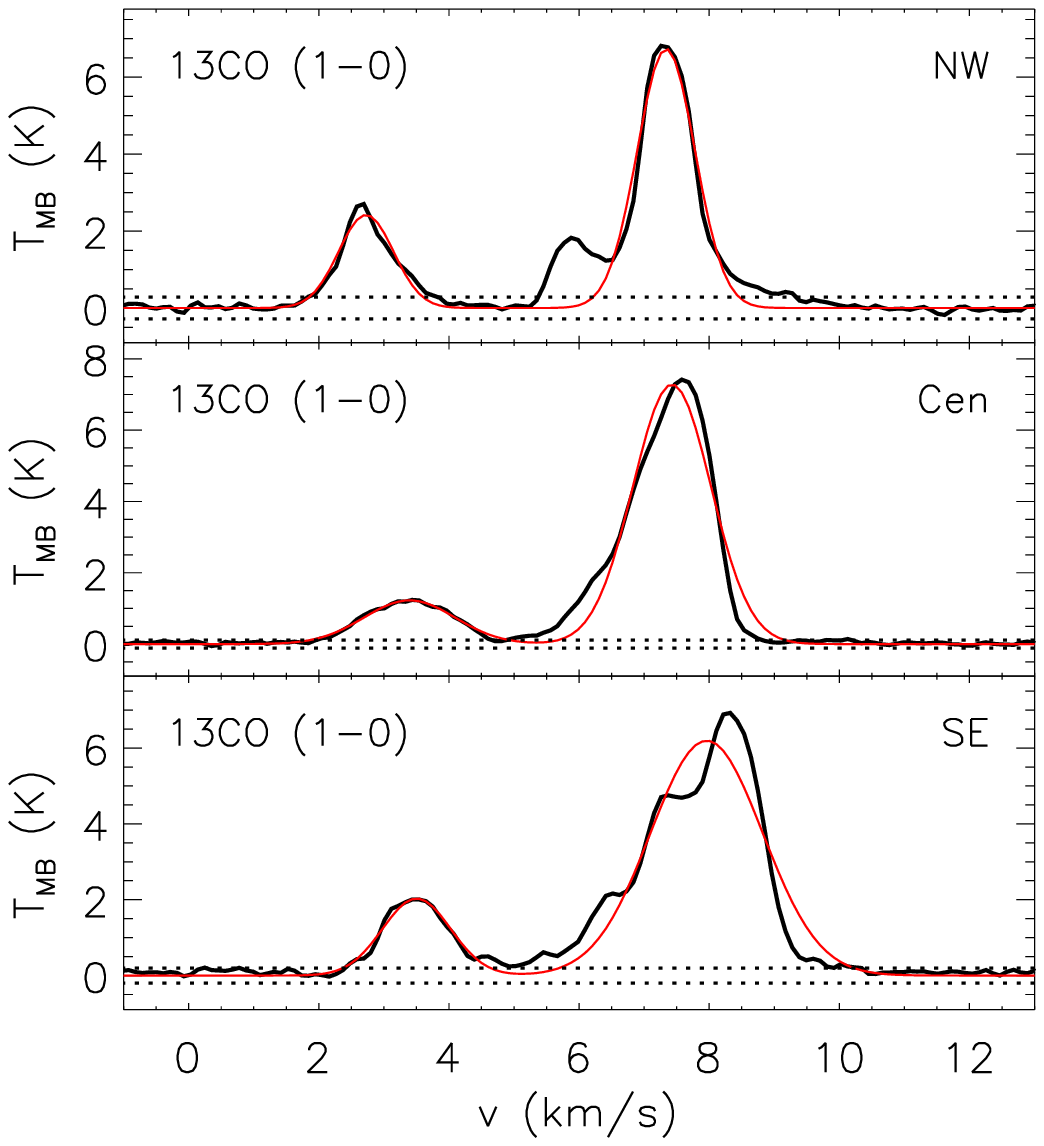} \hspace{0.5mm} \includegraphics[scale=0.75]{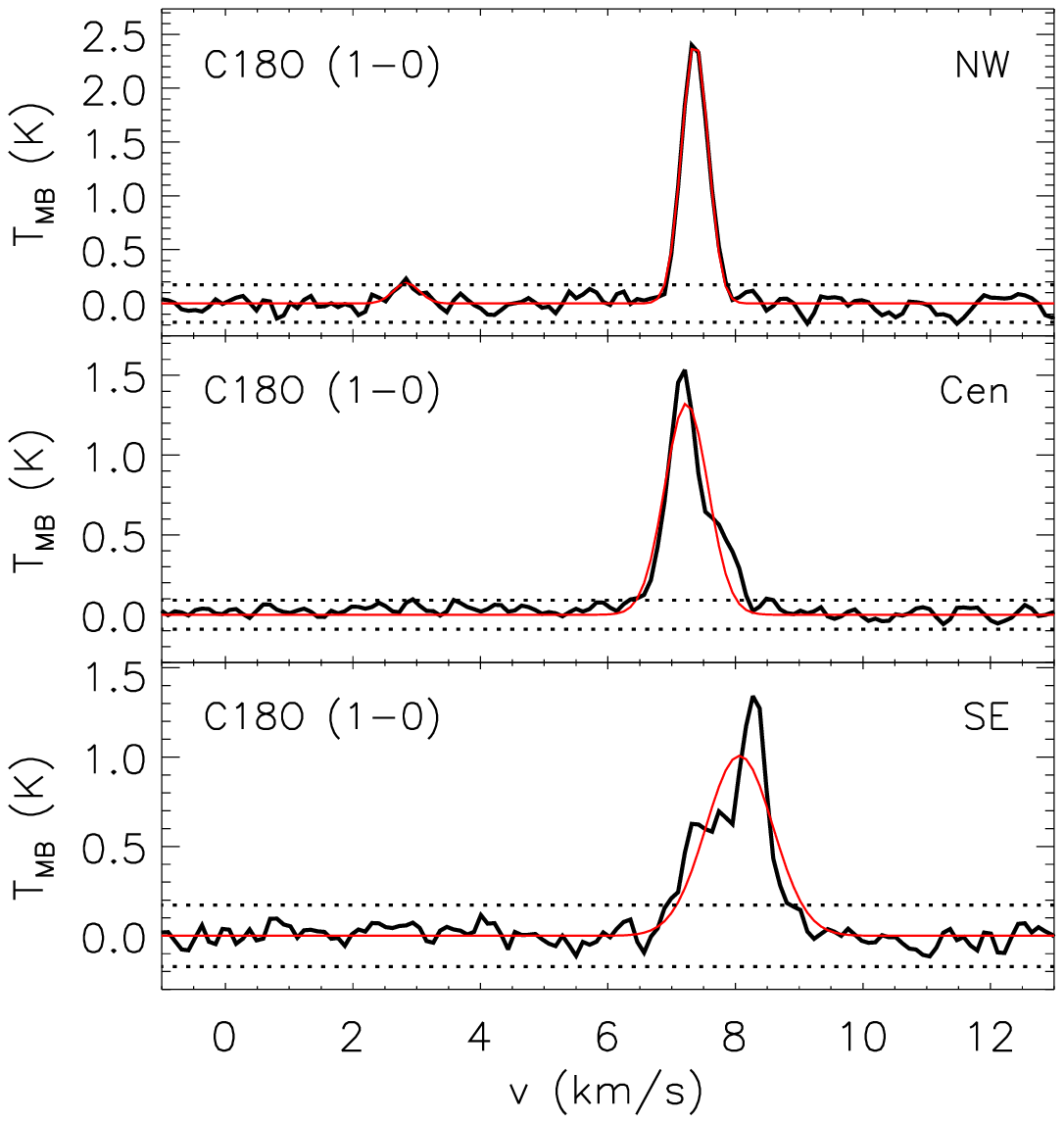}
\caption{Sample IRAM spectra for (left) \CCO\ (1-0) and (right) \COO\ (1-0) towards the north-west corner (NW), center (Cen), and south-east corner (SE) of B1-E.  The pixels used for the spectra are shown in Figure \ref{b1e_colDen}.  The red curves represent the best-fit line profiles using a single Gaussian fit for the 3 \kms\ component and 7.5 \kms\ component, respectively.  The dotted grey lines show the $3\ \sigma$ rms level.  \label{sampleSpec_13co}}
\end{figure}

Figure \ref{sampleSpec_13co} shows three sample IRAM \CCO\ (1-0) and \COO\ (1-0) spectra for the sample pixels toward the north-west (NW) corner, center (C), and south-east (SE) corner of the B1-E clump (see Figure \ref{b1e_colDen}).  The IRAM \CCO\ (1-0) line emission has a main velocity component at $\sim$ 7.5 \kms\ with less prominent emission at $\sim$ 3 \kms.  The integrated intensity associated with the $\sim 7.5$ \kms\ component dominates over the $\sim 3$ \kms\ component by a factor of $\sim 5$ on average.  The \COO\ (1-0) observations barely show any emission at $\sim 3$ \kms, indicating that the gas associated with the $\sim 3$ \kms\ component is more diffuse than the gas found at the $\sim 7.5$ \kms\ (see also, \citealt{Sadavoy12}).  Hereafter, we will only discuss the 7.5 \kms\ velocity component, unless stated otherwise.  

The IRAM \CCO\ (1-0) and IRAM \COO\ (1-0) spectra at $\sim 7.5$ \kms\ reveal very complicated, non-Gaussian profiles indicative of several sub-components at very similar, overlapping velocities.  The best-fit velocity profiles in Figure \ref{sampleSpec_13co} show single Gaussian fits for each of the 3 \kms\ and 7.5 \kms\ components, whereas many IRAM \CCO\ (1-0) spectra require five unique velocity subcomponents to fit well the observed velocity profiles at $\sim 7.5$ \kms, and similarly, the IRAM \COO\ (1-0) spectra need three components at $\sim 7.5$ \kms.  The SMT \CCO\ (2-1), JCMT \CCO\ (3-2), and IRAM \COO\ (2-1) line emission also show complex line profiles.  

\begin{figure}[h!]
\includegraphics[trim=0mm 0mm 0mm 0mm, clip=true,scale=1]{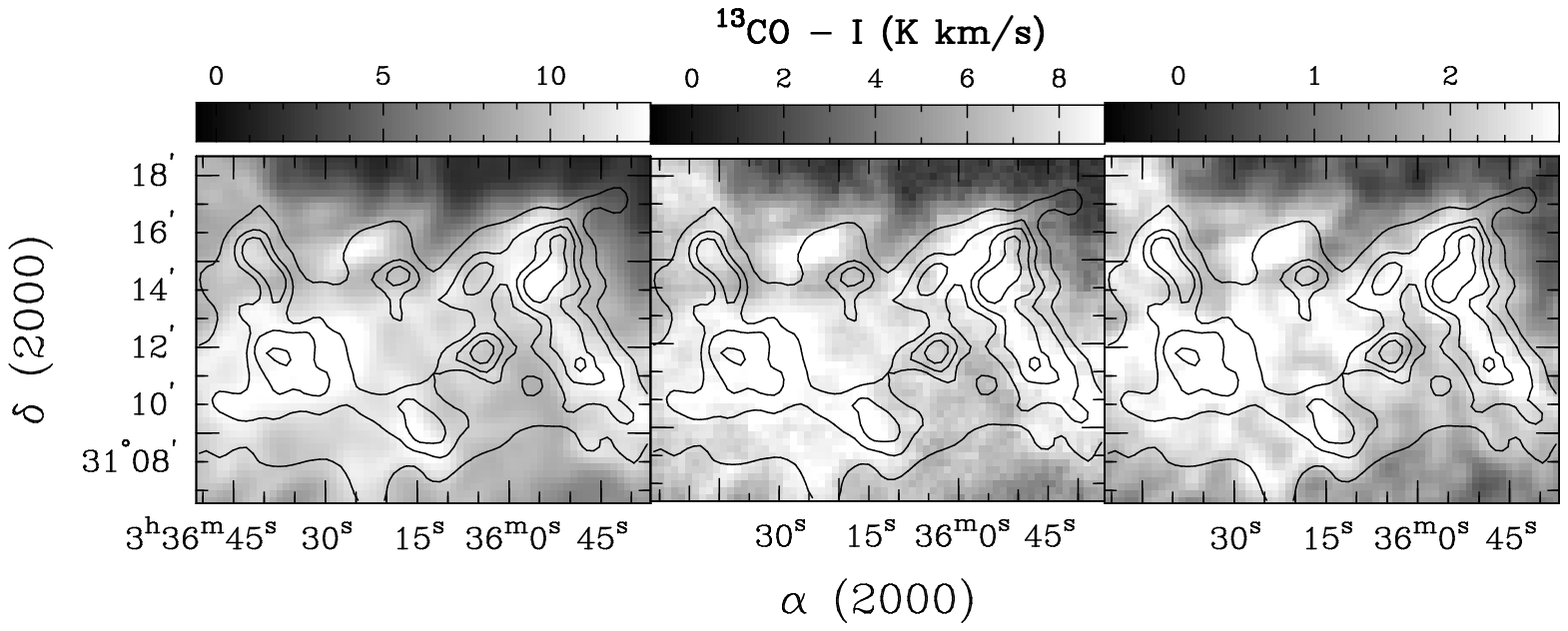}
\caption{Integrated intensity maps for \CCO\ line emission over the velocity range of $5-9$ \kms\ only.  The left panel shows IRAM \CCO\ (1-0), the middle panel shows SMT \CCO\ (2-1), and the right panel shows JCMT \CCO\ (3-2).  The lower greyscale limit corresponds to the 3 $\sigma$ uncertainty in the integrated intensity.  The contours show \emph{Herschel}-derived column density levels of  $5.0 \times 10^{21}\ \cden, 7.0 \times 10^{21}\ \cden, 8.5 \times 10^{21}\ \cden\ \mbox{and}\ 10.5 \times 10^{21}$ \cden\ from the HGBS data (see \citealt{Sadavoy14}).\label{integInt_13co} }
\end{figure}

\begin{figure}[h!]
\includegraphics[scale=1]{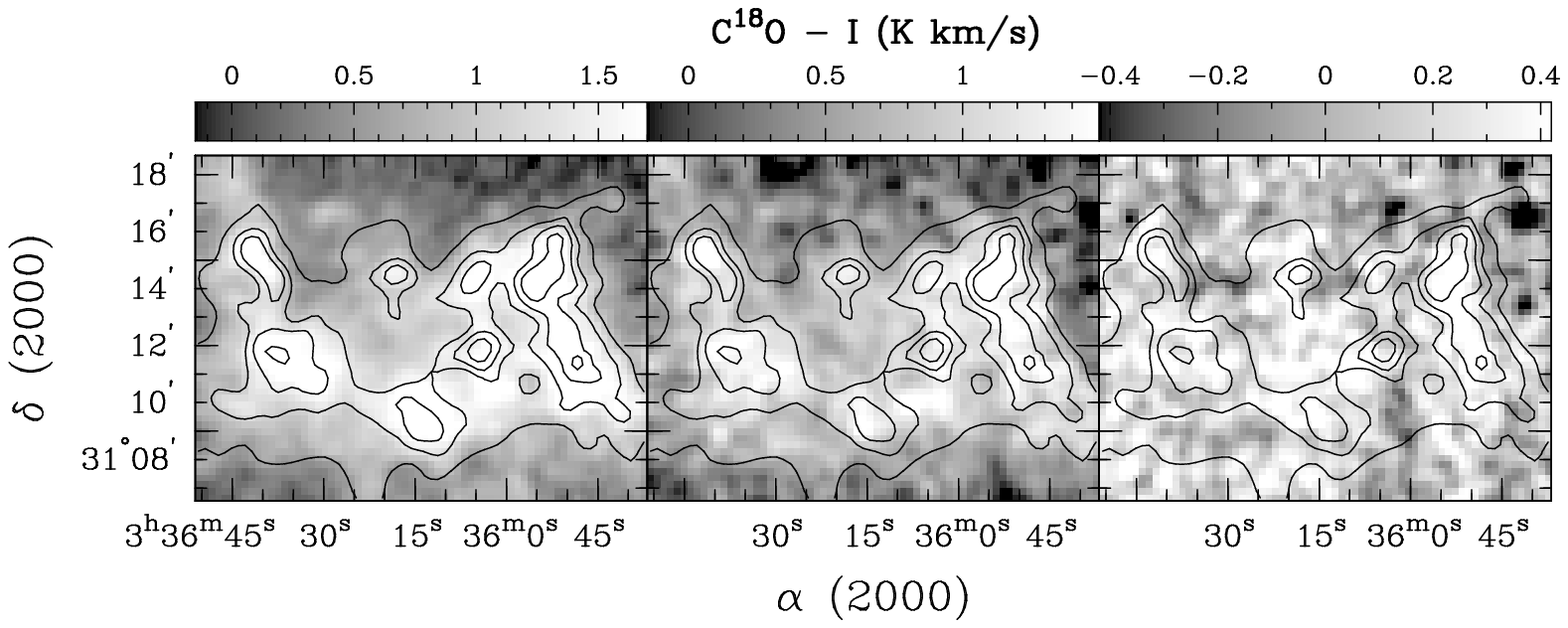}
\caption{Same as Figure \ref{integInt_13co}, but showing the IRAM \COO\ (1-0), IRAM \COO\ (2-1), and JCMT \COO\ (3-2) integrated intensities. \label{integInt_c18o}}
\end{figure}

Figures \ref{integInt_13co} and \ref{integInt_c18o} show respectively the integrated intensity maps for all the \CCO\ and \COO\ transitions across the main velocity structure over $5-9$ \kms.   Both figures include \NHH\ contours to highlight the regions with higher column densities of dust.  We find that the \CCO\ observations show some agreement with the dust, whereas the IRAM \COO\ (1-0) and (2-1) integrated intensities show excellent correspondence with the dust-derived column densities.  Although the right panel of Figure \ref{integInt_c18o} hints at possible JCMT \COO\ (3-2) emission, particularly towards the B1-E1 region (see Figure \ref{b1e_colDen}), any detections are very weak (i.e., $\lesssim 3\ \sigma$).  We do not use the JCMT \COO\ (3-2) data in our subsequent analyses.

\subsection{Velocity Gradients}\label{vel_grad}

\begin{figure}[h!]
\includegraphics[scale=0.51]{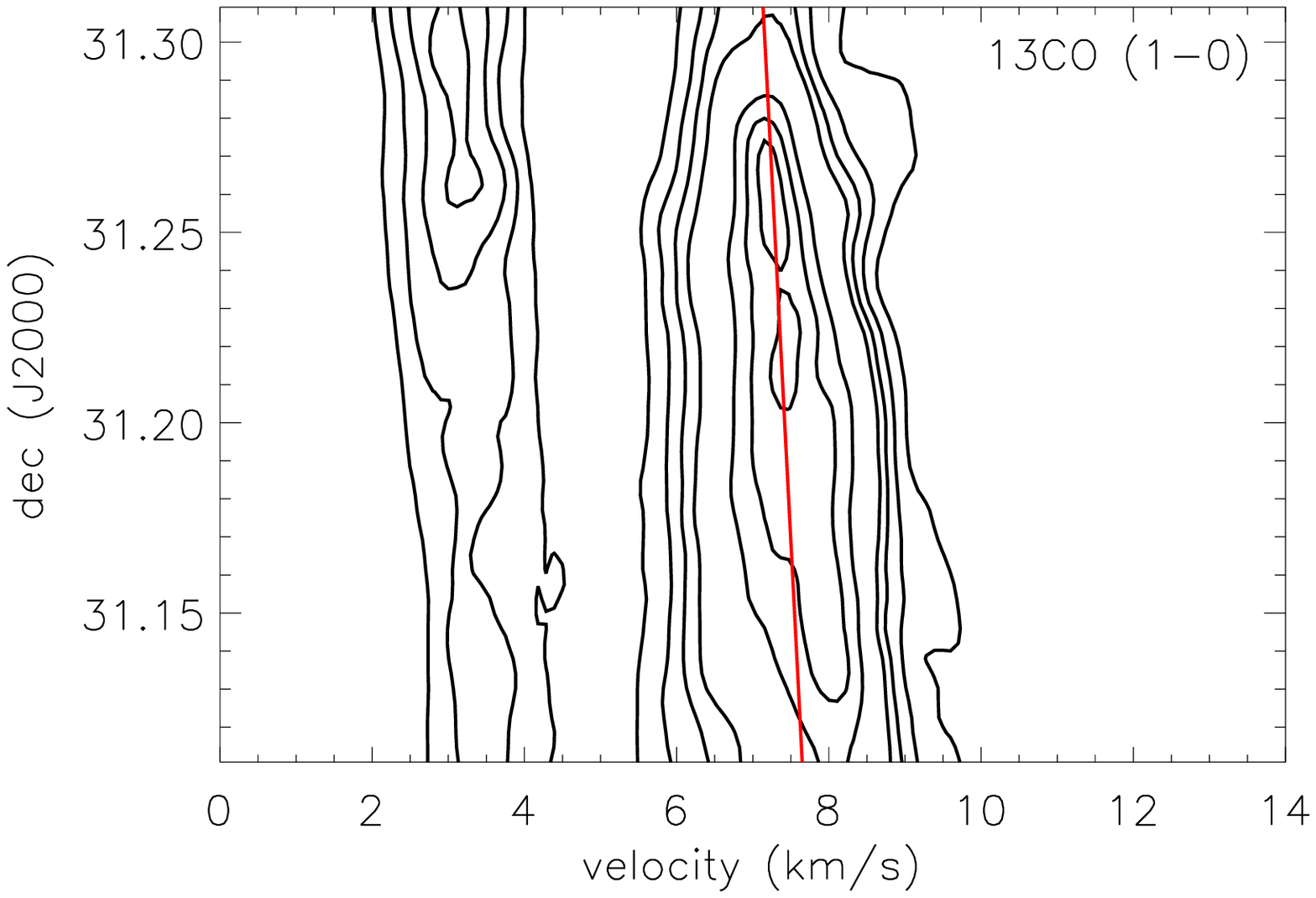} \includegraphics[scale=0.51]{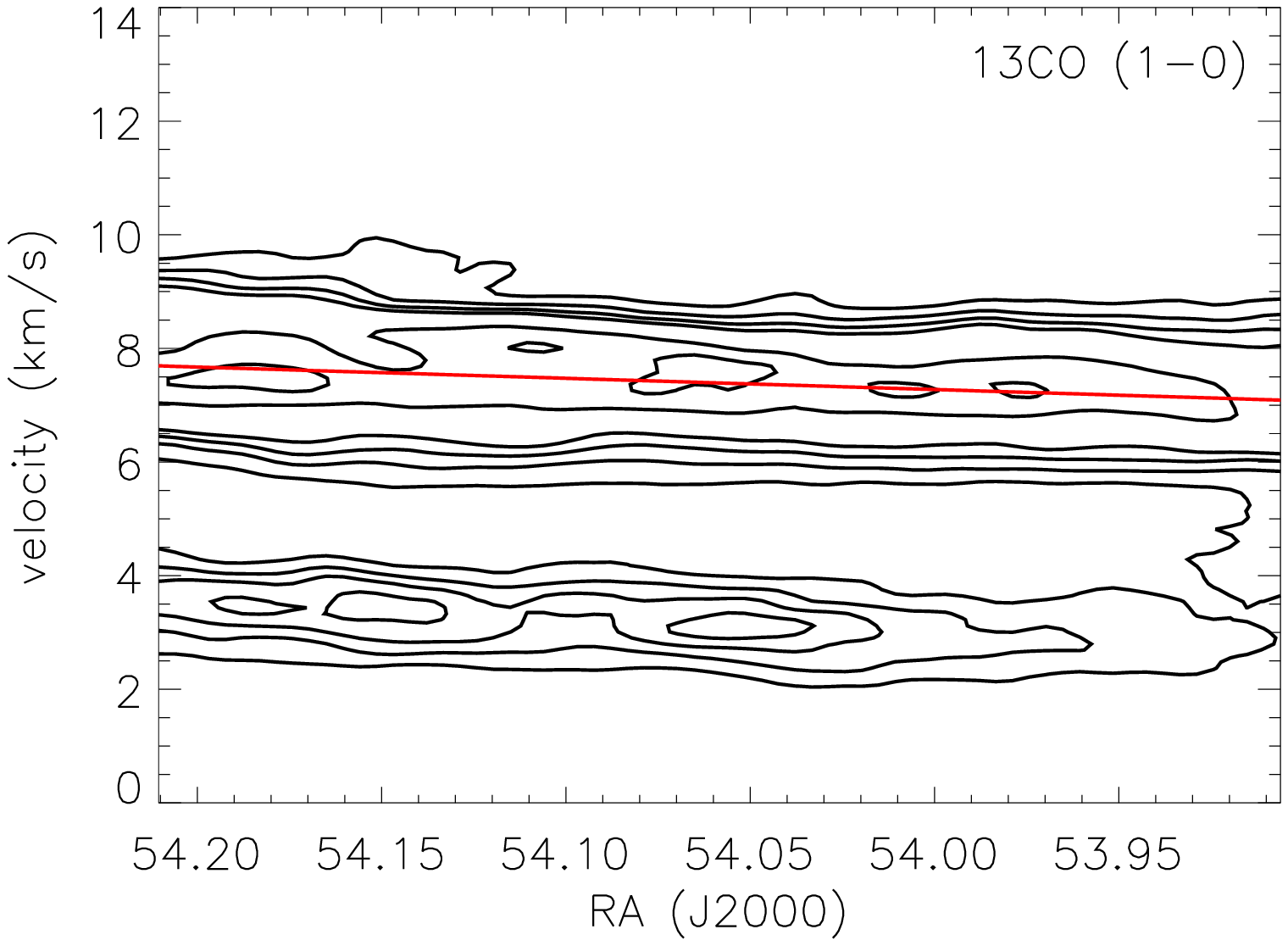} \\[5mm]
\includegraphics[scale=0.51]{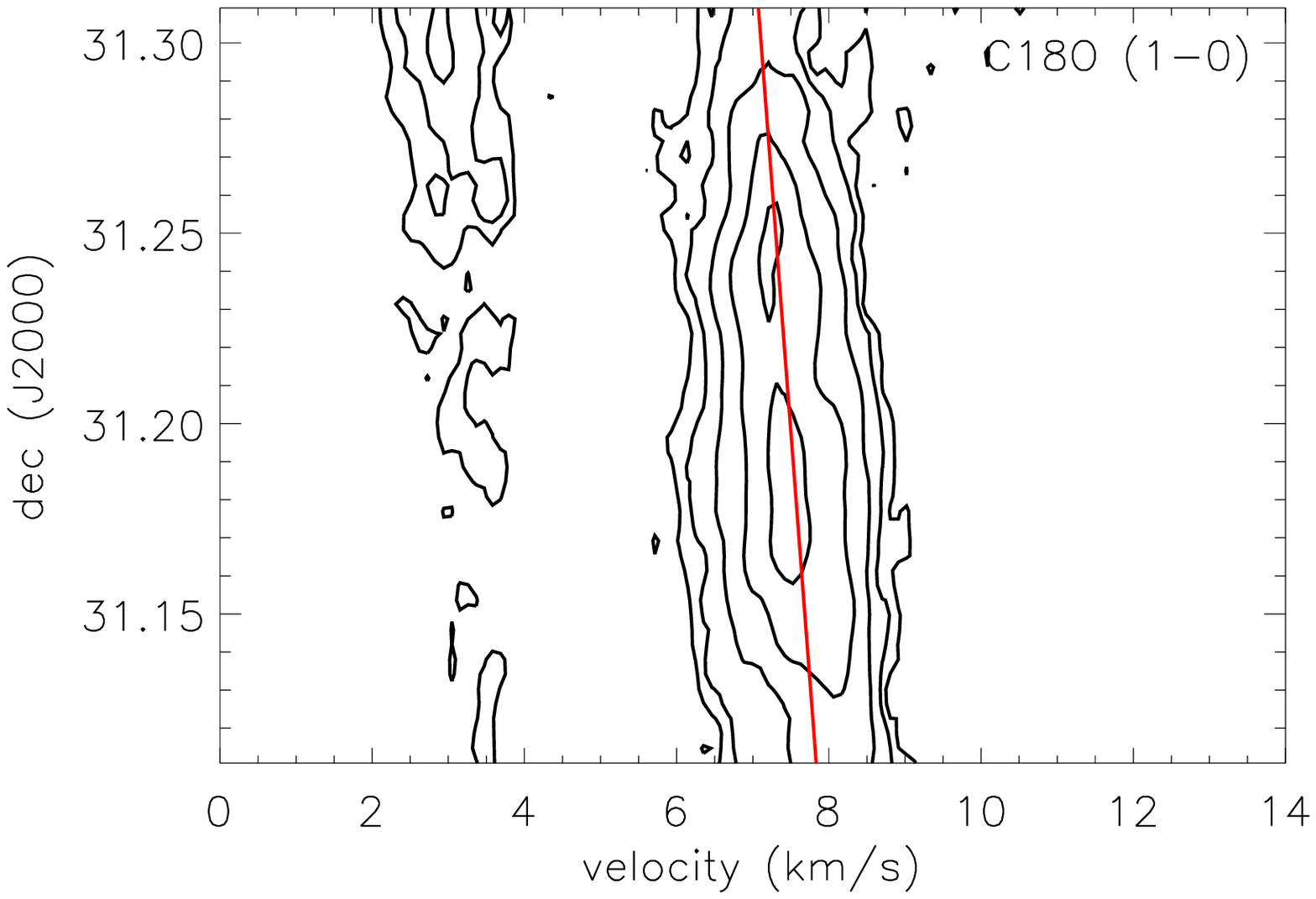} \includegraphics[scale=0.51]{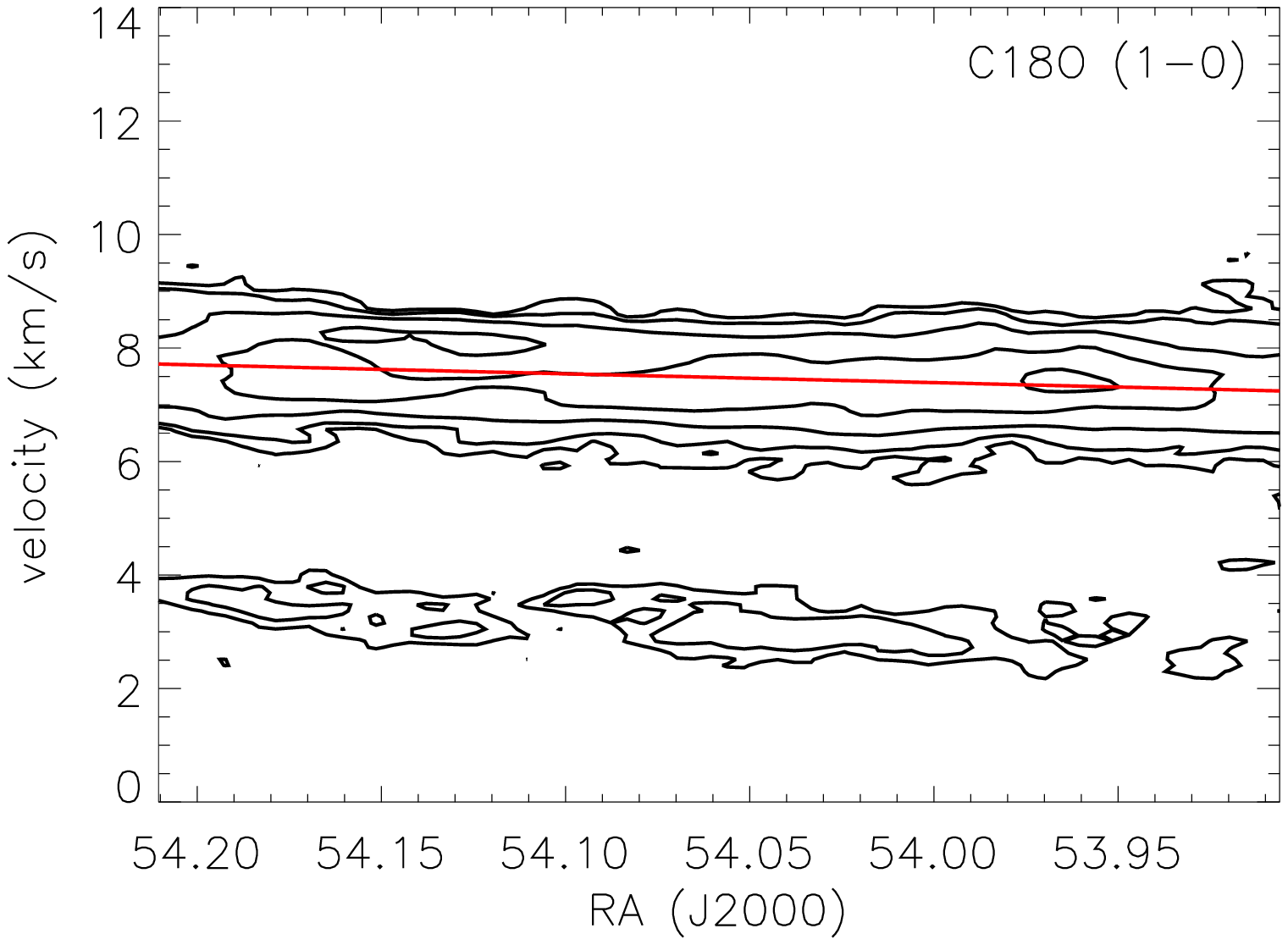}
\caption{Position-velocity diagram for (top) IRAM \CCO\ (1-0) and (bottom) IRAM \COO\ (1-0).   Lines were averaged across strips in declination (left) or Right Ascension (right).  For \CCO\ (1-0), contour levels correspond to average \Tmb\ values of 0.5 K, 1.0 K, 1.5 K, 2.0 K, 4.0 K, 5.0 K, and 5.7 K.  For \COO\ (1-0), contours correspond to average \Tmb\ values of 0.04 K, 0.08 K, 0.2 K, 0.5 K, and 1.0 K.  The red lines correspond to the weighted mean velocity for $\Tmb > 1.0$ K (\CCO) or $\Tmb > 0.1$ K (\COO).}\label{contour}
\end{figure}

The IRAM \CCO\ (1-0) and IRAM \COO\ (1-0) spectra show a radial velocity gradient, where the centroid velocities increase from the NW to the SE  (see Figure \ref{sampleSpec_13co}).   Figure \ref{contour} shows position-velocity diagrams for the IRAM \CCO\ (1-0) and IRAM \COO\ (1-0) line transitions.  We find consistent flux-weighted gradients for both line transitions of 0.94 \kmspc\ from the IRAM \CCO\ (1-0) observations and 1.0 \kmspc\  for the IRAM \COO\ (1-0) observations at an angle of $\sim 20-30$\degree\ north-to-west.  This gradient of $\sim 1$ \kmspc\  is consistent with the results of \citet{Kirk10} for the B1-E clump. We also find a similar gradient and angle with the 3 \kms\ component.   

\begin{figure}[h!]
\includegraphics[scale=1]{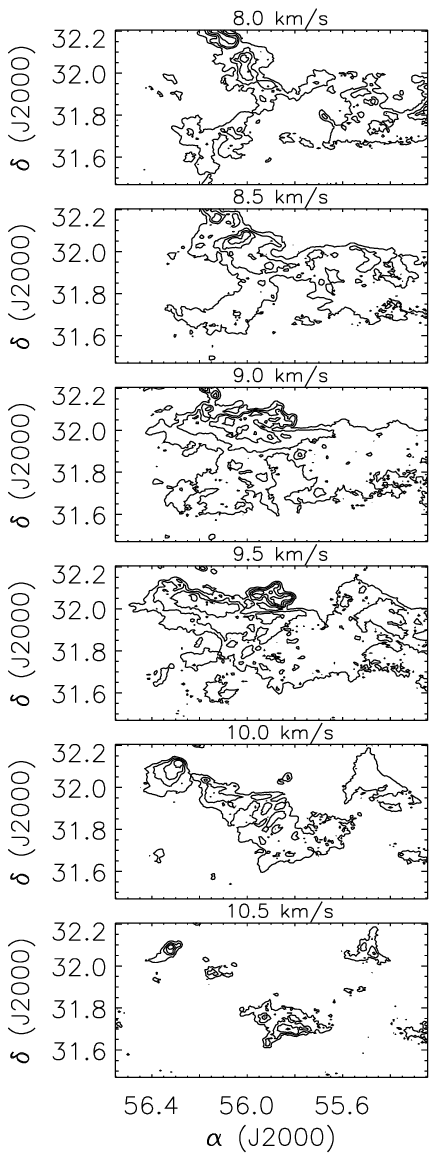} \hspace{7mm} \includegraphics[scale=1]{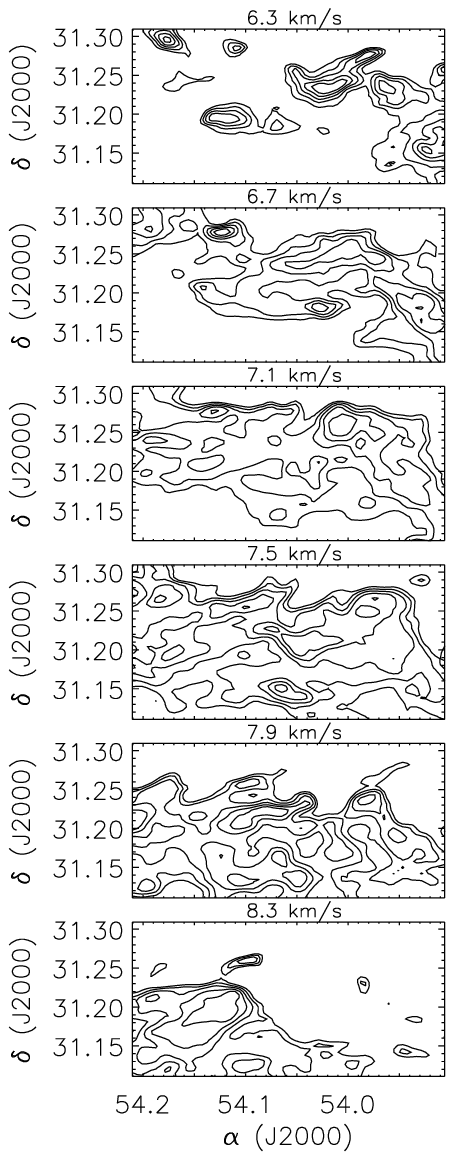} \hspace{7mm}\includegraphics[scale=1]{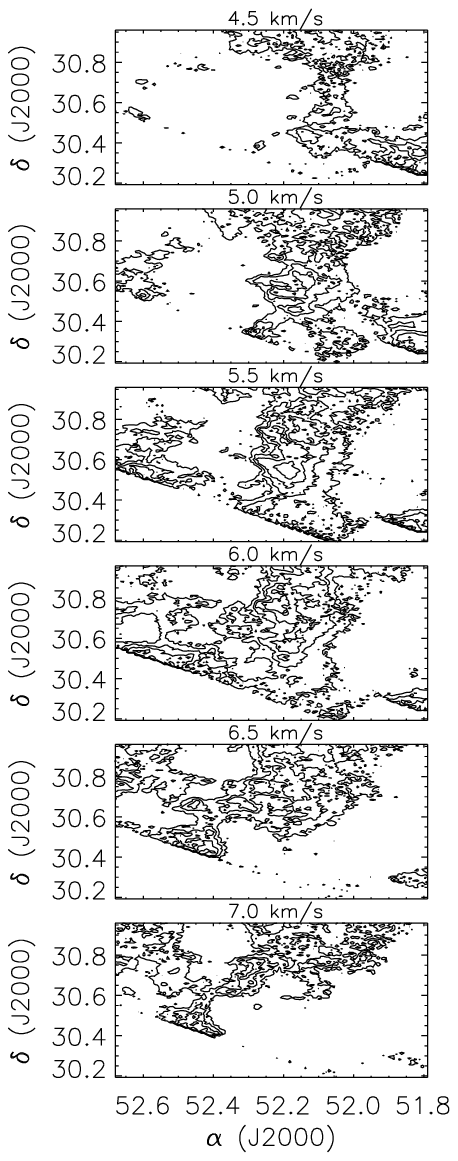}
\caption{Channel maps of \CCO\ (1-0) observations of IC348 (left), B1-E (middle), and L1455 (right).  The IC348 and L1455 maps use data from the COMPLETE survey (\citealt{Ridge06}), whereas the B1-E data use the IRAM \CCO\ (1-0) data presented here.  The contours for IC348 and L1455 correspond to 20\%, 40\%, 60\%, and 80\%\ of the peak line brightness of each channel.  The contours for B1-E correspond to 50-90\%\ of the peak line brightness (in steps of 10\%).  The observed velocity channel is given above each map.}\label{chan_map}
\end{figure}

On much larger scales, a steep velocity gradient across the Perseus cloud has been well identified in previous studies (e.g., \citealt{Ridge06}; \citealt{Sun06}).  Figure \ref{chan_map} compares the channel maps of B1-E to those of L1455 (clump west of B1-E) and IC348 (clump east of B1-E) using data from the COMPLETE survey (\citealt{Ridge06}).  The line-of-sight velocities in Perseus typically increase from the south-west (SW) to the north-east (NE), as illustrated by the channel maps of L1455 and IC348, as well as their centroid velocities.  Moreover, the COMPLETE \CCO\ (1-0) and \CO\ (1-0) data give very similar gradients. 

\begin{figure}[h!]
\includegraphics[scale=1]{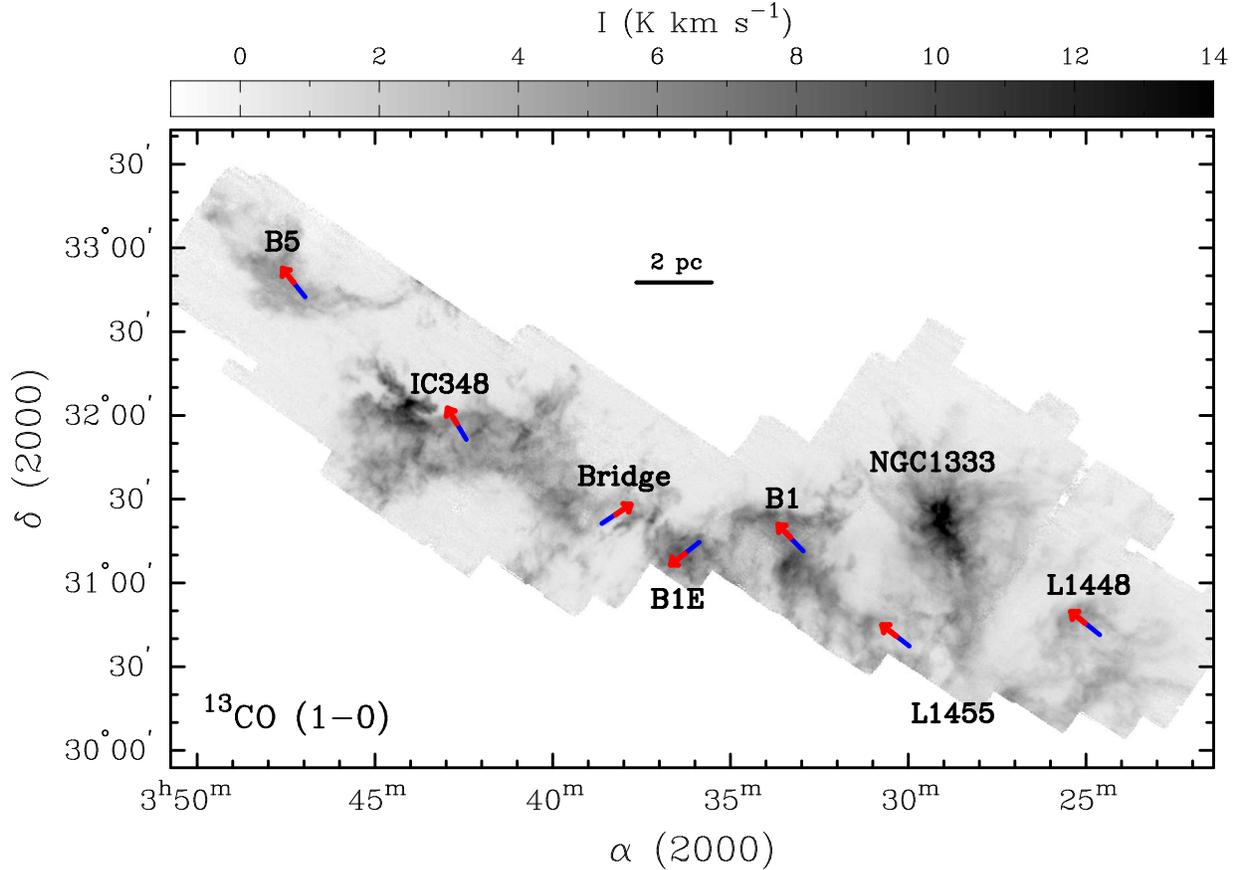} 
\caption{Velocity gradient orientations across Perseus as determined from \CO\ (1-0) observations from COMPLETE (\citealt{Ridge06}).  Background image shows the integrated intensity map for \CCO\ (1-0) using the COMPLETE data.  Arrows show the direction of centroid velocities from bluer values to redder values (they are not scaled to the gradient magnitude).  We excluded NGC1333 from this analysis (see text).}\label{gradient}
\end{figure}

To characterize better the velocity gradients elsewhere in Perseus, we used the \CO\ (1-0) observations from the COMPLETE survey (\citealt{Ridge06}) in a similar manner as our IRAM observations of B1-E.  We broke up the Perseus cloud into rectangular regions identified by-eye to enclose B5, IC348, B1-E, B1, L1455, and L1448 (e.g., see \citealt{Sadavoy14}), as well as additional regions encompassing the ``bridge'' material between IC348 and B1-E (see \citealt{Bally08}).  We excluded NGC1333 due to its large sample of outflows (\citealt{Hatchell07}; \citealt{Arce10}).  Note that we generally selected more material than \citet{Kirk10} in measuring the gradients.  Figure \ref{gradient} shows the radial velocity gradient orientations across Perseus using the same technique outlined in Figure \ref{contour}.  Most regions have gradients of $\sim 0.3-0.6$ \kmspc, where the centroid velocities increase from the SW to the NE at position angles of $\sim 30-50$\degree\ north-to-east.  Thus, the velocity gradients toward B1-E and the nearby ``bridge'' region stand out significantly for increasing instead in directions perpendicular to the velocity gradients found in the other Perseus clumps.  

\subsection{Turbulence}\label{vel_turb}

To quantify the level of turbulent motion in B1-E, we measured the non-thermal velocity dispersion for both IRAM \CCO\ (1-0) and IRAM \COO\ (1-0) line emission using two methods;  (1) we adopted the median velocity dispersion from the single-Gaussian fits (see Section \ref{vel_structure}); (2) we summed all the spectra, where each spectrum was re-centered prior to stacking them (to correct for the observed velocity gradient; see previous section).  Table \ref{turbTable} gives the velocity dispersions associated with each method and each molecule.  For both molecular tracers, the velocity dispersion is slightly higher when using the stacked spectra compared to the median velocity dispersion of all spectra, though both methods agree within uncertainties.    Additionally, our line widths for B1-E agree well with those found in other (more active) clumps of Perseus (e.g., \citealt{Hatchell05}; \citealt{Kirk10}).  

\begin{table}[h!]
\caption{Turbulent Velocity Dispersions Across B1-E}\label{turbTable}
\begin{tabular}{lcc}
\hline\hline
 Line 		& Median\tablenotemark{a} 	&  Stacked Spectrum\tablenotemark{b} 	\\
 \hline
\CCO\ (1-0)	& 0.7 \kms		& 	0.8	\kms		\\
\COO\ (1-0)	& 0.4	 \kms		&	0.5	\kms		\\
\hline
\end{tabular}
\tablenotetext{a}{The distribution of velocity dispersions across B1-E are non-Gaussian.  For the \CCO\ (1-0) and \COO\ (1-0) data at $\NHH\ \ge 5.0 \times 10^{21}$ \cden, $\sim$ 75 \%\ of spectra have velocity dispersions within 0.13 \kms\ and 0.15 \kms\ of the respective median velocity dispersion.}
\tablenotetext{b}{Single Gaussian fits yield robust velocity dispersions with uncertainties better than 1\%.}
\end{table}

We estimated the thermal and non-thermal line contributions using the median dust temperature of $\sim 14$ K, as determined from modified blackbody fitting to the \emph{Herschel} SEDs (see also Appendix \ref{TexAppendix}), as a proxy for the gas kinetic temperature.  Assuming an equivalent gas kinetic temperature of 14 K, the thermal line widths of \CCO\ and \COO\ are $\sim 0.06$ \kms, whereas the observed velocity dispersions are $\sigma \gtrsim 0.3$ \kms.  Although the gas and dust are likely decoupled at the densities the \CCO\ and \COO\ line emission trace, these temperatures should agree within a factor of two (e.g., \citealt{Young04}; \citealt{Ceccarelli07}).  Thus, the expected thermal velocities will be $<0.1$ \kms\ (e.g., for $T_K < 2 T_{dust}$) and the \CCO\ and \COO\ spectra cannot be explained by thermal broadening alone.  Rather, we assume the observed line widths represent the turbulent (non-thermal) velocity dispersion.

\subsection{Substructures}\label{sources}

\begin{figure}[h!]
\includegraphics[scale=0.92]{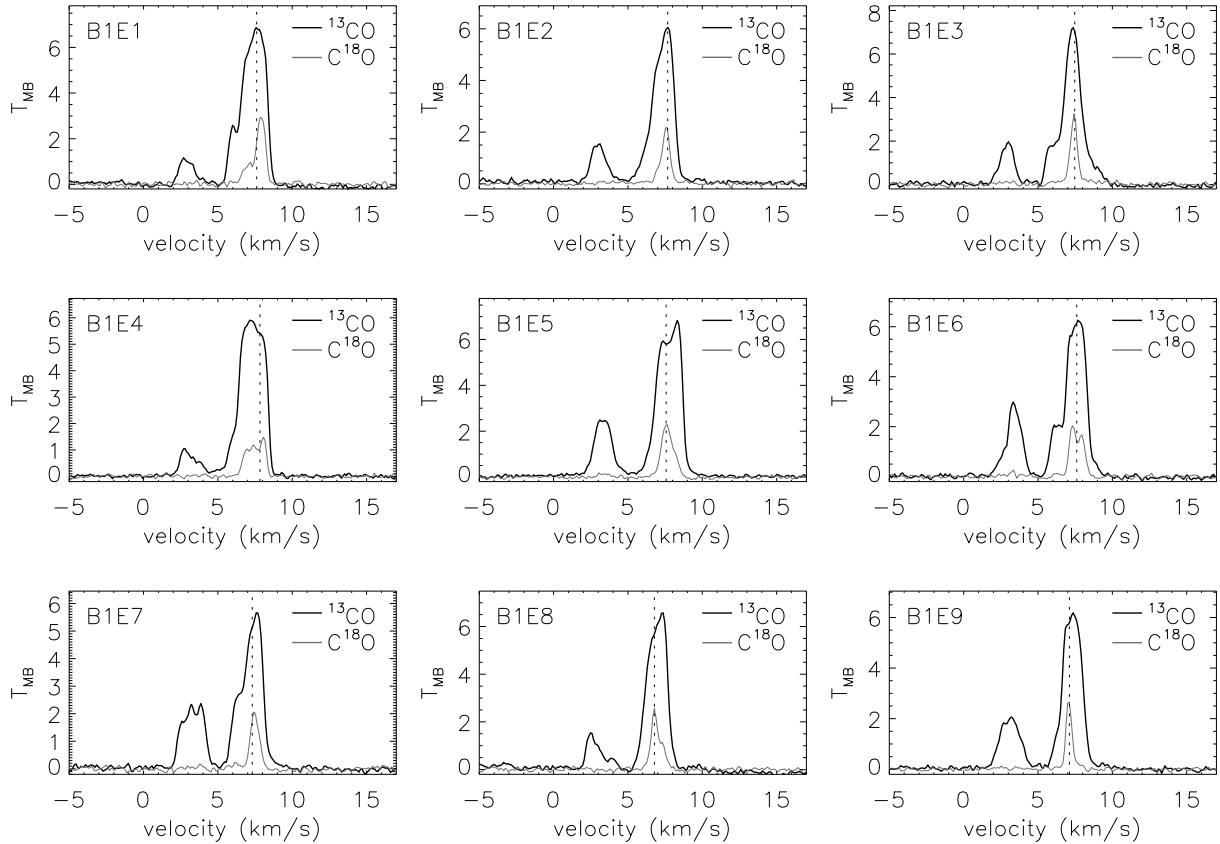} 
\caption{Comparison of the IRAM \CCO\ (1-0) and IRAM \COO\ (1-0) line profiles for the nine densest B1-E substructures.  The dotted vertical line corresponds to the respective \ammonia\ (1,1) centroid velocity from \citet{Sadavoy12}.}\label{substructures}
\end{figure}

\begin{figure}[h!]
\includegraphics[scale=0.92]{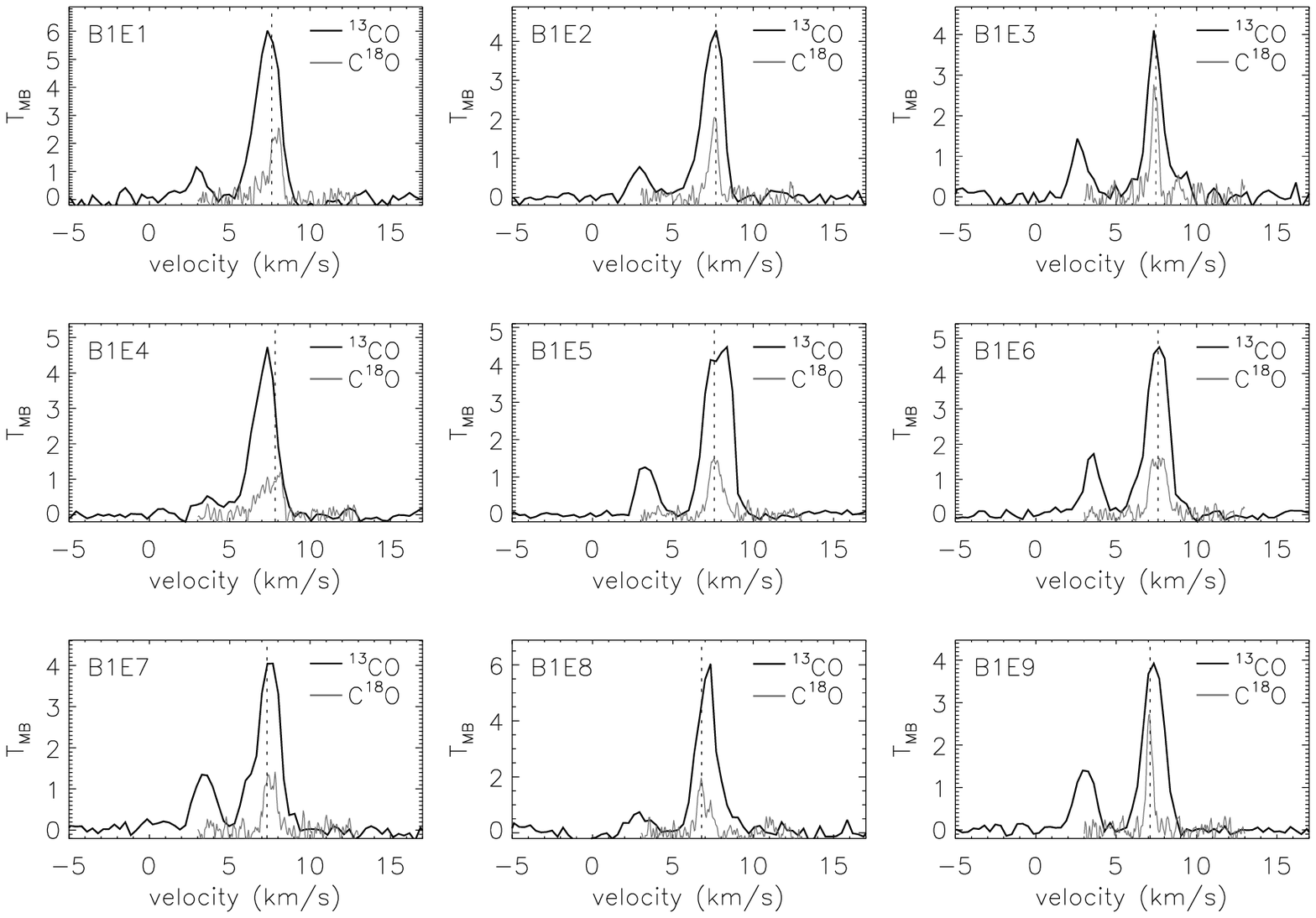} 
\caption{Same as Figure \ref{substructures} but for the SMT \CCO\ (2-1) and IRAM \COO\ (2-1) transitions.}\label{substructures21}
\end{figure}

\citet{Sadavoy12} identified nine substructures in B1-E (see also, Figure \ref{b1e_colDen}) and characterized them with \ammonia\ (1,1) emission.   Figures \ref{substructures} and \ref{substructures21} compare the (1-0) and (2-1) spectra of both \CCO\ and \COO, respectively, with the \ammonia\ (1,1) centroid velocities (dashed line) of these nine objects.  The \ammonia\ (1,1) data trace the denser central material of each substructure, whereas the \COO\ and \CCO\ data trace better the source envelopes and more diffuse clump, respectively.  In general, most substructures show a good agreement between the \ammonia\ (1,1) centroid velocities and the \COO\ velocity peaks, in agreement with previous studies (e.g., \citealt{Walsh04}; \citealt{Kirk07}; \citealt{Andre07}; \citealt{Kirk10}).  This similarity supports the notion that the dense material within cores forms quiescently within their envelopes (see also, \citealt{Offner08}).  The \CCO\ spectra show more deviations in centroid velocity (and higher line widths), which is unsurprising as \CCO\ is a better tracer of lower density ($\sim 10^3$ \vol) gas.

The B1-E1 and B1-E4 substructures are the only objects with significant ($> 3\ \sigma$) deviations between the peak-line velocities of \COO\ and \ammonia\ (Figures \ref{substructures} and \ref{substructures21}).   In particular, these objects have asymmetric \COO\ profiles with a narrow Gaussian-like peak and a broad blue-shifted ``knee,'' where the \ammonia\ emission peaks between the Gaussian-peak and the ``knee.''  In addition to these, the B1-E6 source has a double peaked \COO\ spectrum (a feature seen in both transitions) with single-Gaussian \ammonia\ (1,1) emission at velocities directly between the two \COO\ peaks.   Normally, a double-peaked spectrum with brighter emission toward the bluer side would suggest self-absorption and perhaps infall.  Nevertheless, \COO\ emission is generally optically thin (see Section \ref{radex}) and as such, will not produce such a feature.  Moreover, the \CCO\ spectra, which should be more optically thick, do not show this double-peaked profile.  Thus, the B1-E6 spectra may correspond to multiple velocity subcomponents along the line of sight.  For example, the bluer \COO\ velocity peak matches the centroid velocity of the nearby B1-E7 source, whereas the redder peak appears associated with a separate velocity structure east of the B1-E6 and B1-E7 ridge.  We discuss the possibility of multiple subcomponents in Section \ref{discussion}.

In addition to similar core-envelope velocities, several studies have compared the core-to-core centroid velocity dispersion with the turbulence in their surrounding environment.  From \citet{Sadavoy12}, the substructures have an \ammonia-derived core-to-core relative motion of $\sigma_{sys} \sim 0.24$ \kms.  This core-to-core centroid velocity dispersion matches well the dispersion of the \COO\ (1-0) peak velocities ($\sim 0.29$ \kms), but is factors of 2-3 lower than the typical velocity dispersions observed across the entire clump (see Table \ref{turbTable}).  \citet{Kirk10} similarly found that the starless (and protostellar) cores in other regions of Perseus systematically had much lower relative motions than the turbulence detected in the more diffuse, ambient clump material (see also, \citealt{Kirk09}).  Therefore, these core precursors do not appear to move in a manner similar to the B1-E clump as a whole, but rather, they have more homogenous velocities like those seen in more evolved dense cores.  We also find that the \ammonia\ detections show no indication of a velocity gradient, whereas the \COO\ data clearly show a $\sim 1$ \kmspc\ radial velocity gradient across B1-E.  Thus, the B1-E substructures appear to be decoupled (i.e., following a distinctly different velocity distribution) from the moderately dense gas as traced by the \COO\ emission, suggesting that the systemic motions of dense cores somehow homogenize very quickly in their evolution (see also, Section \ref{compSim}).  

The B1-E substructures have very broad \ammonia\ emission compared to typical observations of dense gas in prestellar cores.  For example, \citet{Rosolowsky08} conducted a targeted \ammonia\ survey of Perseus, and from their results, the prestellar cores have \ammonia\ velocity dispersions $\sigma < 0.25$ \kms, with an median value of $\sigma \sim 0.15$ \kms.   In comparison, most of the B1-E substructures have much broader \ammonia\ velocity dispersions of $\sigma \gtrsim 0.3$ \kms\ (\citealt{Sadavoy12}).  The most line-broadened substructures have similar velocity dispersions ($\sim 0.4$ \kms) as the larger-scale \COO\ gas.   If these broad line widths represent the turbulent motions within these objects, the B1-E substructures have significant internal pressures.  Assuming a typical substructure density of $10^4$ \vol\ (see \citealt{Sadavoy12}), the substructures have internal pressures of $P_{sub}/k = \rho\sigma^2 = (0.3 - 6) \times 10^5$ \Kcm.   In comparison, the volume-averaged gravitational pressure of the B1-E clump can be estimated from \citet{BertoldiMckee92} as
\begin{equation}
\frac{\bar{P}}{k} = 1.01\ a_1 \left<\phi_G\right> \left(\frac{M}{\Msun}\right)^2 \left(\frac{R}{1 \mbox{ pc}}\right)^{-4} \Kcm,
\end{equation}
where $M$ is the clump mass, $R$ is the observed clump size, $a_1$ is a scaling factor corresponding to a nonuniform density distribution, and $\left<\phi_G\right>$ is a scaling factor corresponding to the cloud geometry ($\phi_G = 1$ for a perfect sphere).  Following \citet{Lada08}, we assume $a_1=1.3$, appropriate for a self-gravitating cloud.  We also adopt $\left<\phi_G\right> \sim 1.01$, since B1-E is fairly circular ($\sim 1.2$ aspect ratio; see \citealt{BertoldiMckee92}).  Thus, the B1-E volume-average gravitational pressure is $\sim 3 \times 10^5$ \Kcm, which is very similar to the internal pressures for the B1-E substructures.  Moreover, if the substructures are centrally concentrated, their surface pressures may be much lower than their average internal pressures (\citealt{Lada08}), suggesting that the external pressure of the clump is sufficient to keep the B1-E substructures confined.  Such pressure-confined structures have been seen more recently in different environments, including Taurus (\citealt{Seo15}) and Ophiuchus (\citealt{Pattle15}).

\section{Radiative Transfer Models}\label{radex}

We used the radiative transfer code, \emph{RADEX}, by \citet{vanDerTak07} to characterize the abundances of \CCO\ and \COO.  In brief, \emph{RADEX} is a non-LTE radiative transfer code that models molecular transitions to determine the expected line strengths, optical depths, and energy level populations for different transitions of a given molecule.  The code assumes a homogenous, spherical cloud with a user-determined temperature and abundance.   Since our spectra contain multiple velocity components, we used the results from our single-Gaussian fits (see Figure \ref{sampleSpec_13co}) to determine the observed line brightnesses and widths for all our transitions.  Thus, our \emph{RADEX} results represent the bulk properties of the B1-E clump.

We ran \emph{RADEX} for individual pixels across our \CCO\ and \COO\ observations.  (Recall that all data have been convolved to a common resolution and rebinned to a common grid.)  We selected only those pixels with $\NHH \ge 5 \times 10^{21}$ \cden, to ensure that we had strong detections $> 5\ \sigma$ for all transitions considered.  For each pixel, we adopted a background temperature of 2.73 K and the line widths from our single Gaussian fits.  Additionally, we assumed the kinetic gas temperature is equal to the observed dust temperature from SED fitting of the \emph{Herschel} data (see \citealt{Sadavoy14}).  Since the B1-E is a moderately dense region, the gas and dust temperatures may not be coupled.  We expect differences of only a few Kelvin, which produce relatively minor uncertainties in the models (see Appendix \ref{TexAppendix}).

We applied 2500 combinations for the H$_2$ number density, \nHH, and the molecular column density, \NCCO\ or \NCOO, in our analysis.  For \nHH, we adopted values between $10^3\ \vol\ \le \nHH\ \le 10^5\ \vol$\ in equal spacings of $\Delta\log[\nHH] = 0.04$.  For the molecular column densities, we adopted $5 \times 10^{14}\ \cden\ \le \NCCO\ \le 5 \times 10^{16}\ \cden$\ and $10^{14}\ \cden\ \le \NCOO\ \le 10^{16}\ \cden$\ with equal spacings of $\Delta\log[\mbox{N(X)}] = 0.04$.  For each density/column density pair, we compared the model line brightnesses and the observed line brightnesses for all observed transitions, and we adopted the parameters associated with the best-fit model.

\begin{figure}[h!]
\includegraphics[trim=0mm 0mm 0mm 0mm,clip=true,scale=0.575]{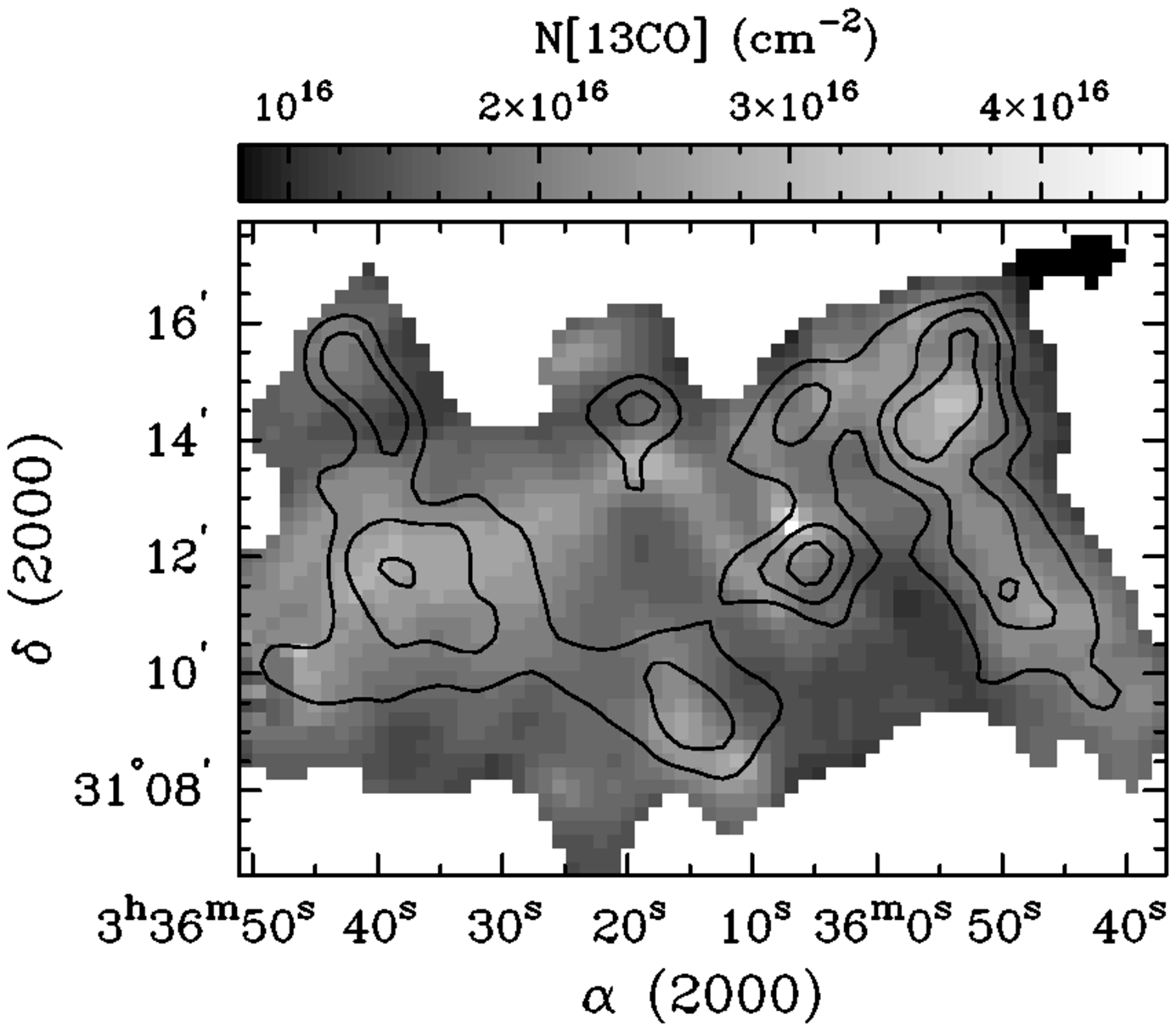} \hspace{0mm}
\includegraphics[trim=0mm 0mm 0mm 0mm,clip=true,scale=0.575]{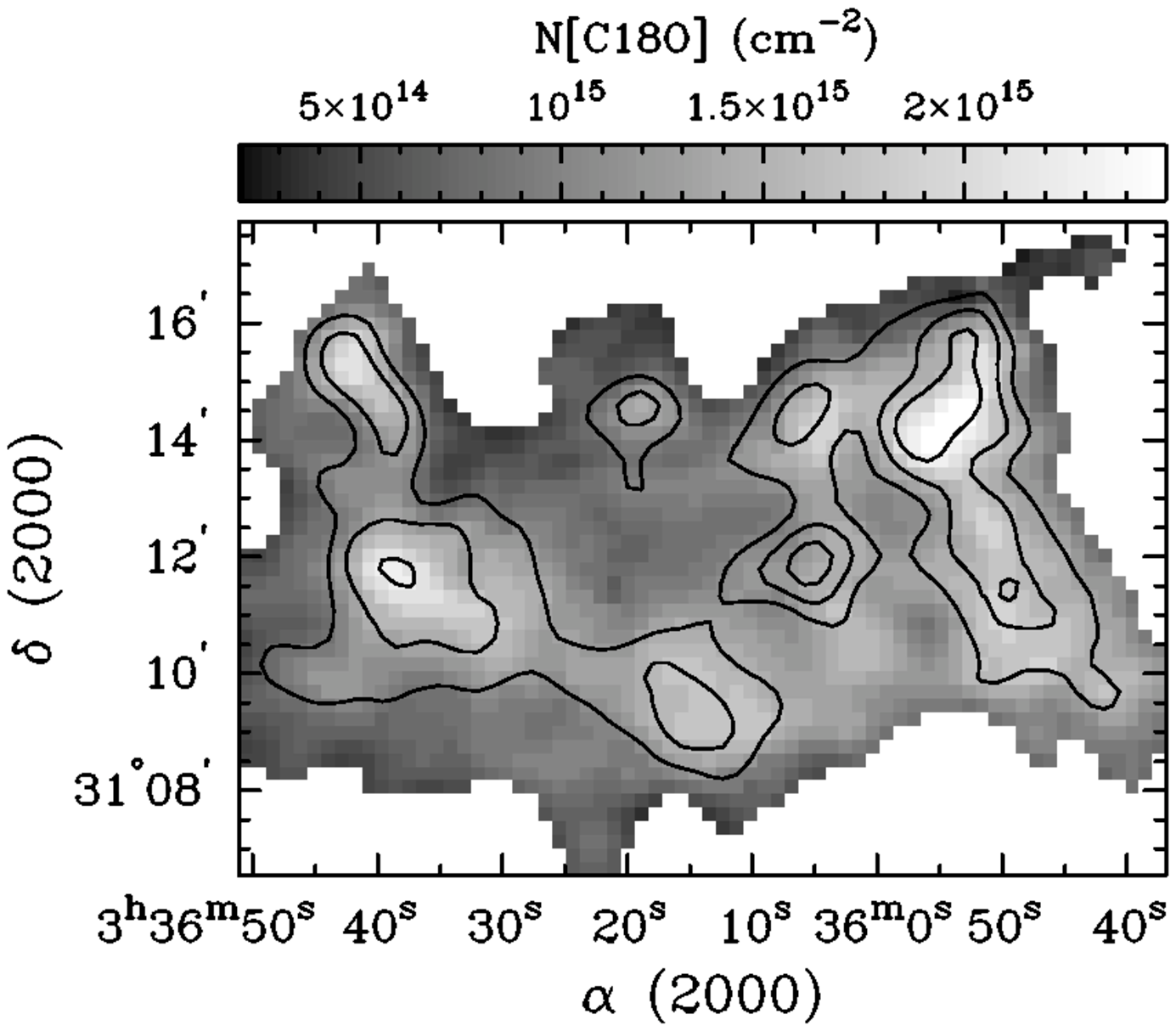}\\[2mm]
\includegraphics[trim=0mm 0mm 0mm 0mm, clip=true, scale=0.575]{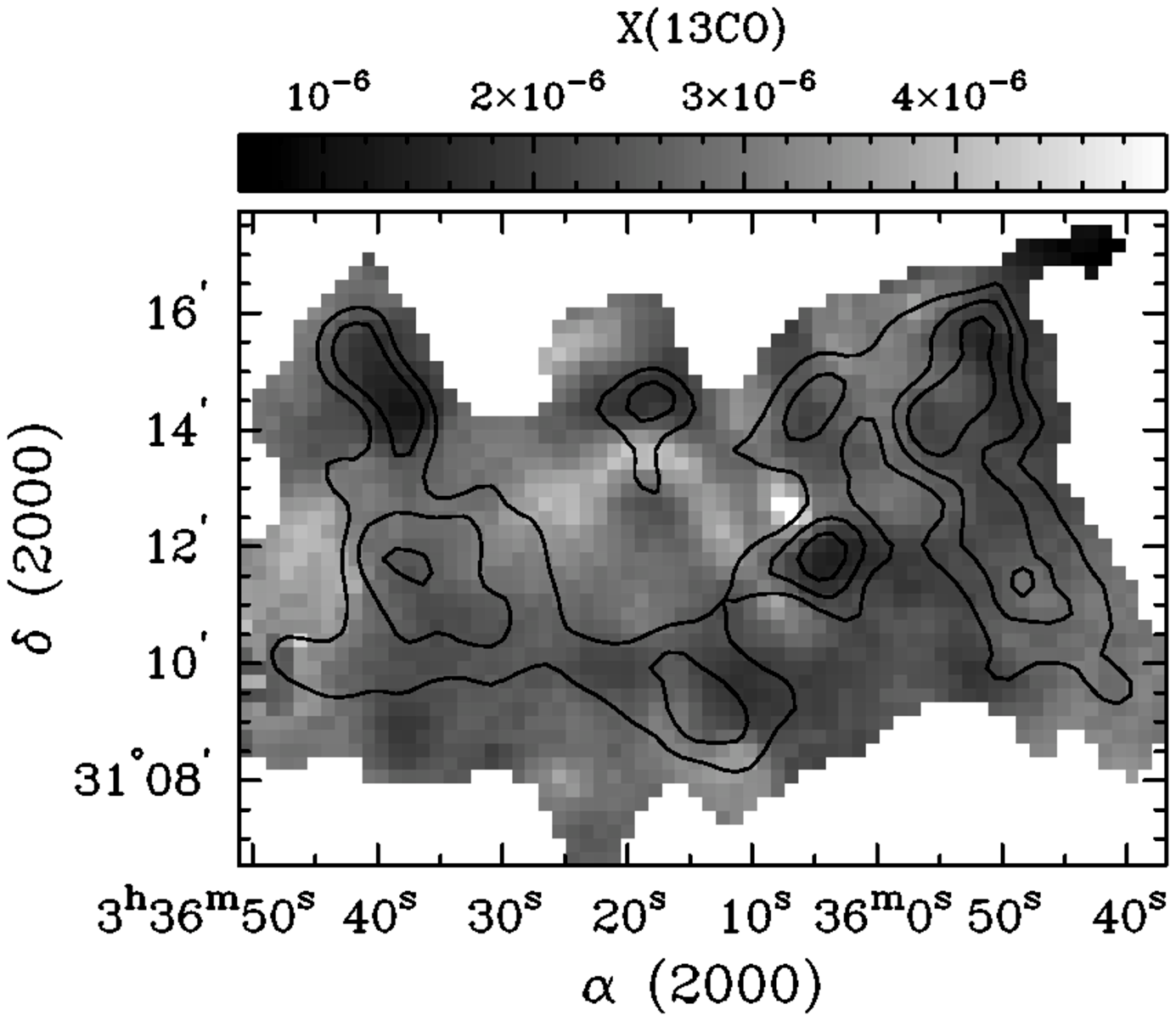} \hspace{0mm}
\includegraphics[trim=0mm 0mm 0mm 0mm, clip=true,scale=0.575]{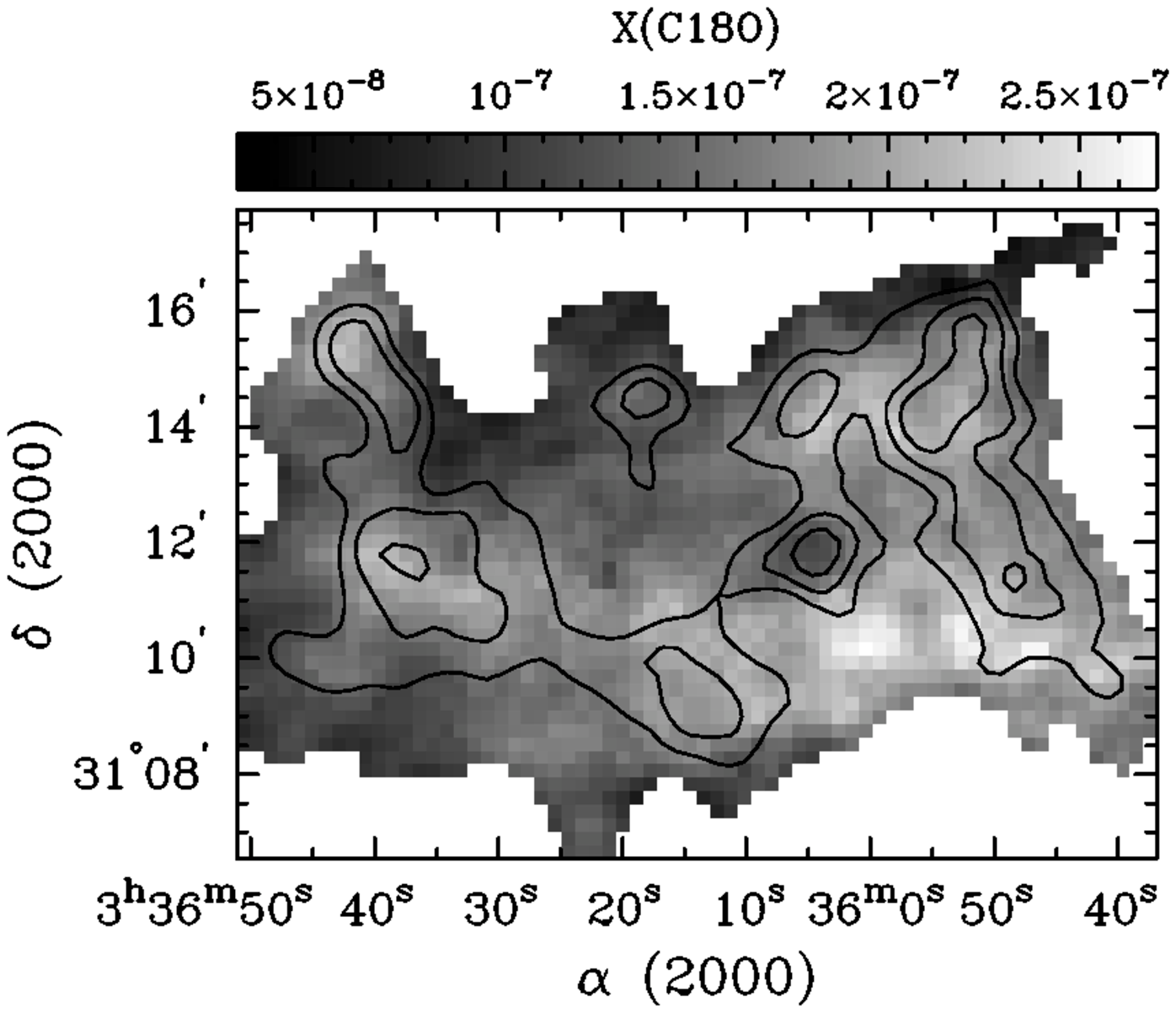}
\caption{\emph{Top:} Column density maps of (left) \CCO\ and (right) \COO\ from our best-fit RADEX models. \emph{Bottom:} Ratios of (left) \CCO\ and (right) \COO\ column densities with the \emph{Herschel}-derived H$_2$ column density from the HGBS data.   Contours correspond to $\NHH = 7.0 \times 10^{21}\ \cden, 8.5 \times 10^{21}\ \cden,$ and $10.5 \times 10^{21}$ \cden. }\label{RADEX_colDen_maps}
\end{figure}

Figure \ref{RADEX_colDen_maps} shows the best-fit column densities for \CCO\ and \COO\ from the RADEX models as well as their corresponding X-factors, using our \emph{Herschel}-derived H$_2$ column densities (see Figure \ref{b1e_colDen}) to determine N(X)/\NHH\ for both \CCO\ and \COO.  We find that the \COO\ column densities trace the densest material (see contours) very well, whereas the \CCO\ column densities do not follow the dense material as closely.  For the \CCO\ emission, we find moderate optical depths of $\sim 3$ at all J-lines connecting the material around B1-E2 to B1-E8 and B1-E9.  On average, the $J = 2-1$ transition has the highest optical depths, whereas the $J= 1-0$ has the lowest optical depths.  In contrast, the \COO\ emission is optically thin, suggesting that we are tracing all the emission along the line of sight.    

\begin{figure}[h!]
\includegraphics[scale=0.8]{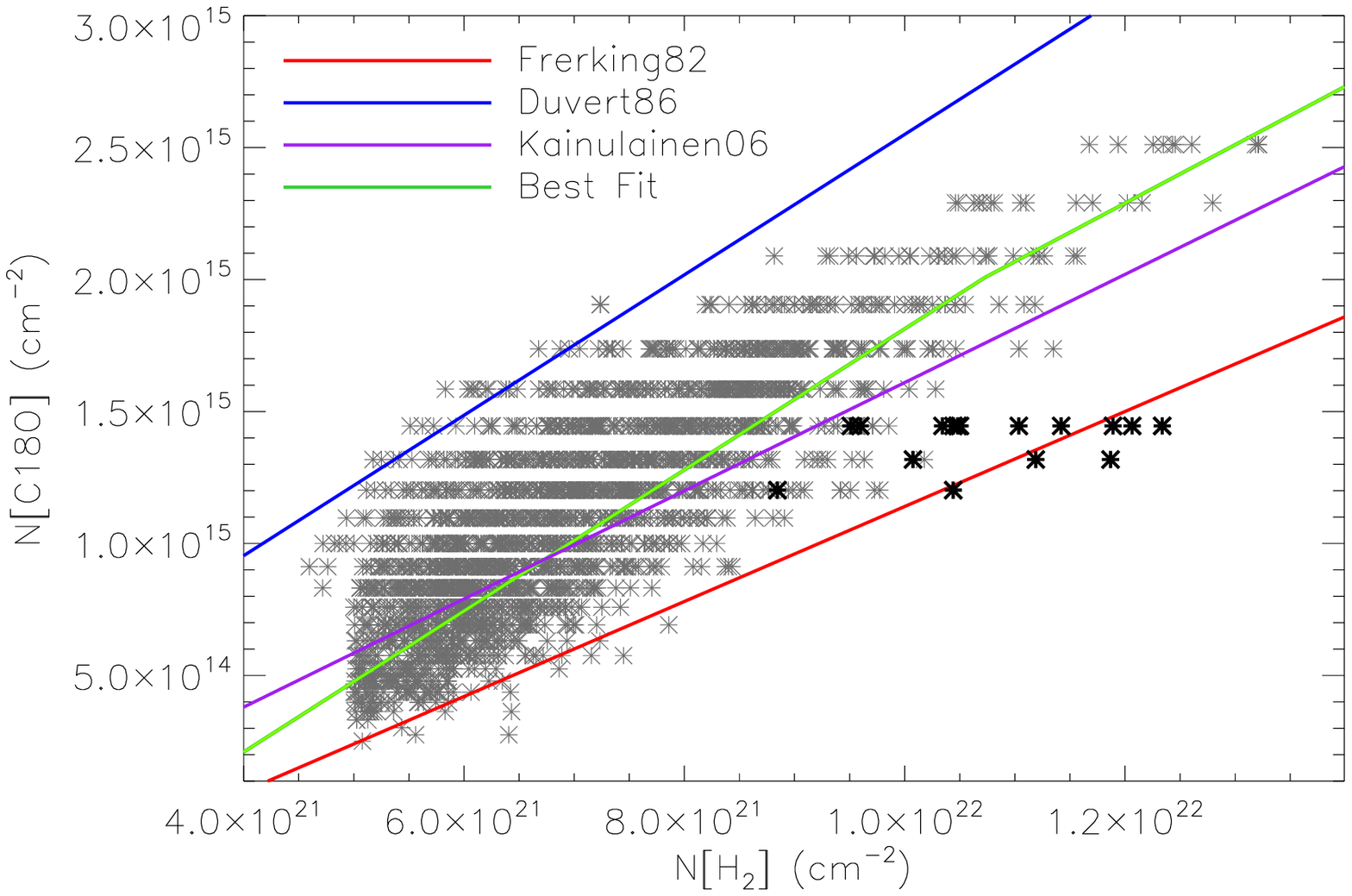} 
\caption{Comparison between N(\COO) and \NHH\ for B1-E.  The black points represent the column densities associated with B1-E2.  The red curve shows the \XCOO\ relation for Ophiuchus (for $\Av > 4$ mag) from \citet{Frerking82}, the blue curve shows the \XCOO\ relation for Taurus from \citet{Duvert86}, and the purple curve shows the average \XCOO\ for Chameleon from \citet{Kainulainen06}.  Since all three studies primarily fit data with lower \Av\ values (e.g., $\Av < 7$ mag), we used the \NHH/\Av\ relation from \citet{Bohlin78} of $\NHH/\Av = 9.4 \times 10^{20}$ \cden\ mag$^{-1}$.  The Bohlin et al. ratio better represents lower column density environments (e.g., $\NHH < 6 \times 10^{21}$ \cden) than the relation from \citet{Draine03} of $\NHH/\Av = 6.9 \times 10^{20}$ \cden\ mag$^{-1}$ (\citealt{Roy14}).   The green curve corresponds to our best-fit relation (excluding the B1-E2 data points) from a two-component linear least-squares regression.  We find a best-fit break in the curve at $\NHH = (10.7 \pm 0.2) \times 10^{21}$ \cden, with $\XCOO\ \sim  2.7 \times 10^{-7}$ at lower column densities than the break and $\XCOO\ \sim 2.2 \times 10^{-7}$ at higher column densities.}\label{C18O_colDen_comp}
\end{figure}

Figure \ref{C18O_colDen_comp} compares the observed values of N(\COO) and \NHH\ with the \XCOO\ relations from \citet[][for Ophiuchus]{Frerking82}\footnote{We use only the high-\Av\ relation of Frerking et al. to better match our B1-E data with $\Av > 7$ mag.}, \citet[][for Taurus]{Duvert86}, and \citet[][for Chameleon]{Kainulainen06}.  The Frerking et al. relation generally underestimates the \COO\ abundances and the Duvert et al. relation generally overestimates them.   The Kainulainen et al. relation agrees well with N(\COO) at low column densities (e.g., $\NHH < 8 \times 10^{21}$ \cden), but underestimates the \COO\ abundances at higher column densities.  Moreover, for $\NHH\ \lesssim 8.0 \times 10^{21}$ \cden, we find that the \COO\ column densities are much lower, on average than what is expected by the \XCOO\ relations.  These lower \COO\ abundances could be due to selective dissociation of \COO\ molecules, which depends greatly on the self-shielding of the molecule (abundance) and geometry of the clump.  Indeed, \citet{vanDishoeck88} showed that the photodissociation rate of \COO\ in the centers of clouds is generally much higher than the photodissociation rate of \CO\ and \CCO\ due to self-shielding.   At the very highest column densities (\NHH\ $\gtrsim 1.0 \times 10^{22}$ \cden), there is far less scatter in the \COO\ abundances, suggesting that the \COO\ gas is more shielded. 

To account for selective photodissociation better, we used a linear least-squares regression to fit simultaneously two distinct linear relations to the lower-density material ($\NHH \lesssim 1 \times 10^{22}\ \cden$) and to the higher density material ($\NHH \gtrsim 1 \times 10^{22}\ \cden$), excluding the data points associated with B1-E2 (black symbols).  We find a best-fit ``break'' in the two relations (e.g., where \COO\ becomes well shielded) at $\NHH = (10.7 \pm 0.2) \times 10^{21}$ \cden\ ($\Av \sim 15$ mag), with $\XCOO \sim 2.7 \times 10^{-7}$ at column densities below this break value and $\XCOO \sim 2.2 \times 10^{-7}$ at higher column densities.  Similarly, both \citet{Frerking82} and \citet{Kainulainen06} found that the N(\COO) relation tends to flatten at higher \Av\ values, though the break identified by Frerking et al. occurs at a column density below what is considered here for B1-E, e.g., at $\Av \sim 4$ mag, whereas we only use data at an equivalent of $\Av > 7$ mag.  Therefore, \COO\ self-shielding may start at $\Av \sim 4$ mag, resulting in a first break in \XCOO, and then, depending on the cloud geometry, the \COO\ gas is mostly self-shielded by $\Av \sim 15$ mag.   We note that this higher \Av\ break for well-shielded gas is similar to the $\Av \sim 12$ mag break \citet{Kainulainen06} found for Chameleon.

Our best-ft relation for \XCOO\ excluded the column densities associated with B1-E2, which appear systematically lower than all the other B1-E substructures (see Figure \ref{C18O_colDen_comp}).  This deviation could indicate that the B1-E2 source has different dust properties than observed elsewhere in B1-E.  For example, if the true dust opacity toward B1-E2 is higher by a factor of two (e.g., due to dust coagulation, see \citealt{Ossenkopf94}), then its \XCOO\ ratios would match the rest of the substructures.  Nevertheless, to produce such a significant difference in the local dust opacity toward just B1-E2, the physical properties of this object must be significantly different than the other substructures.  For example, \citet{Ossenkopf94} find that densities that varied by two orders of magnitude will have dust opacities (from coagulation alone) that differ by less than a factor of two in the \emph{Herschel} bands.  Since B1-E2 has a similar density and dust temperature as the rest of the substructures (e.g., \citealt{Sadavoy12}; Pezzuto et al. in preparation), we would not expect a correspondingly large increase in dust opacity.  

In addition to the lower \COO\ abundances, B1-E2 has an enriched chemistry of nitrogen-bearing molecules and carbon-chain molecules compared to the other substructures (see \citealt{Sadavoy12}), indicative of a distinct chemistry toward only this substructure.  One explanation for this distinct chemistry is that the carbon-bearing molecules are more depleted from the gas phase for B1-E2 and that the lower \XCOO\ factors for B1-E2 are due to this freeze-out of \COO\ molecules.  Many studies have shown lower abundances of \COO\ toward the centers of prestellar cores or filaments (e.g., \citealt{Bacmann00}; \citealt{DuarteCabral10}; \citealt{Hernandez11}) and \citet{Bergin02} found a flattened, slightly decreasing profile in both \COO\ and C$^{17}$O emission toward very high column densities in B68, in good agreement with our flattened N(\COO) profile for B1-E2.  We discuss \COO\ depletion from molecular freeze-out in more detail in Section \ref{compObs}.

\section{Discussion}\label{discussion}

\subsection{Formation of High Density Objects}

The formation mechanism for the B1-E substructures must explain why the B1-E clump is only condensing to form cores now, whereas the other clumps in Perseus have formed populations of dense cores and YSOs (e.g., \citealt{Kirk06}; \citealt{Evans09}).  One simple explanation could be that the condensation of the B1-E clump itself was triggered only recently.  B1-E has a radial velocity gradient nearly perpendicular (in the plane of the sky) to the typical velocity gradient associated with the main Perseus cloud (see Section \ref{vel_grad}).  Thus, this clump may be impacted by large-scale processes such as shearing flows or collisions that have only occurred recently.  Indeed, we see unusual, non-Gaussian \COO\ spectra toward B1-E1 and B1-E4 that match well the spectra associated with colliding clouds modeled by \citet{KetoLattanzio89}.  Since these two sources are located along the western-edge of B1-E at the junction between the B1-E velocity gradient and the Perseus overall gradient, they are most likely to be affected by the interaction of these gas flows (see Section \ref{sources}).  Nevertheless, large-scale flows are generally associated with subsonic motions and filamentary morphologies (e.g., \citealt{Klessen00}; \citealt{Bate03}; \citealt{LiNakamura04}; \citealt{ClarkBonnell05}; \citealt{Klessen05}; \citealt{Basu09}), and the B1-E clump and its substructures shows no evidence of these properties.

The relationship between the B1-E velocity gradient and that of the Perseus cloud remains unclear.  For example, perpendicular velocity gradients may arise from (1) independent clouds along the line of sight, (2) material flowing into or out of the system, (3) dynamic turbulence on small scales (e.g., \citealt{Smith12}), or (4) strong magnetic fields.  At this time, we cannot distinguish between these possibilities to explain the perpendicular velocity gradient between B1-E and the Perseus cloud.  Specific predictive simulations of clump fragmentation triggered by low-velocity shearing (or colliding) flows are needed to determine the origin and impact of the perpendicular velocity gradient in B1-E.    

If the condensation of the B1-E clump was not triggered by an external process, its star formation could be delayed compared to the other Perseus clumps.  \citet{Sadavoy12} suggested that B1-E may be under the influence of a strong, localized magnetic field.  B1-E has uniquely strong R-band polarization vectors (\citealt{Goodman90}), and a strong magnetic field has been shown to both delay star formation and to form objects in a loose configuration without filaments as seen in B1-E (e.g., \citealt{LiNakamura04}; \citealt{Basu09}).  Alternatively, B1-E and its substructures may be influenced by significant turbulent processes (see Sections \ref{vel_structure} and \ref{sources}).  Indeed, \citet{Pon14} found excess emission in high-J \CO\ lines toward the B1-E5 substructure, which they interpreted as evidence of low-velocity shocked gas.    Such small-scale turbulent shocks are expected to form few sources in isolated morphologies on longer time scales than large-scale, high velocity processes (e.g., \citealt{Klessen00}). Thus, the loose configuration of the B1-E substructures and their delayed formation with respect to the other Perseus clumps may indicate that this region is dominated by small-scale turbulent shocks or magnetic fields.

\subsection{Evolution of High Density Objects}

\subsubsection{Comparisons with More Evolved Perseus Clumps}\label{compObs}

The B1-E clump and its substructures have some similarities and differences compared to other, more evolved regions of Perseus.  For example,  we find typical line widths of $\sigma \sim 0.8$ \kms\ and $\sigma \sim 0.4$ \kms\ from the IRAM \CCO\ (1-0) and IRAM \COO\ (1-0) observations, respectively, in excellent agreement with the equivalent turbulent results for the Perseus cloud in general from \citet{Kirk10} and \citet{Hatchell05}.   Moreover, the B1-E substructures appear decoupled from their parent clump.  They do not follow the velocity gradient seen in the larger-scale \CCO\ and \COO\ emission and their central densest material, as traced by \ammonia\ emission (from \citealt{Sadavoy12}), have more homogenous velocities compared to the moderately dense gas traced by either \CCO\ or \COO.  Similarly, \citet{Kirk10} found that the dense (prestellar) cores in the more evolved regions of Perseus also had more homogenous centroid velocities compared to the surrounding, larger-scale gas.   Therefore, B1-E has global kinematic properties similar to those of the other Perseus clumps. 

In contrast, the B1-E substructures were not well detected by SCUBA or Bolocam, whereas both instruments were able to detect well prestellar cores in other regions of Perseus.  Additionally, the B1-E objects have broad, supersonic \ammonia\ line widths ($\sigma \gtrsim 0.3$ \kms), whereas the general prestellar core population in Perseus are more quiescent ($\sigma < 0.25$ \kms; \citealt{Rosolowsky08}).  Such broad line widths could indicate either significant internal turbulent motions or multiple, unresolved subcomponents within the GBT beam (e.g., see \citealt{Smith12}; \citealt{Bailey15}).  Since the B1-E clump is not a very dense region and \ammonia\ (1,1) is not well detected off the column density peaks (see \citealt{Rosolowsky08}), line-of-sight coincidences of unrelated subcomponents of dense gas are unlikely to explain the broad \ammonia\ (1,1) emission for all the substructures (except possibly B1-E6, where we detect two distinct velocity components in the \COO\ line emission). Instead, the central dense gas itself may be fragmented into different velocity subcomponents. Without higher resolution observations, however, we cannot identify multiple sources.

One of the B1-E substructures stands out as unique.  B1-E2 is the only object with transonic \ammonia\ (1,1) line widths ($\sigma \sim 0.1$ \kms), similar to what is detected in typical prestellar cores.  Moreover, B1-E2 also has relatively lower \COO\ column densities than the other substructures for the same column densities (see Figure \ref{C18O_colDen_comp}), which could indicate \COO\ depletion uniquely toward this object, another property associated with prestellar cores.    This decrease in \COO\ abundance is well within the uncertainties of \XCOO, however.  For example, \NHH\ is uncertain within a factor of 2 from assuming a single dust opacity (e.g., see \citealt{Ossenkopf94}).  Moreover, our RADEX models assume a single system temperature, single system velocity, and a spherical geometry which may not accurately represent the B1-E conditions.  (Note that our dust opacity assumptions produce more significant uncertainties than adopting the dust temperature as a proxy of the kinetic gas temperature; see Appendix \ref{TexAppendix}.) Nevertheless, B1-E2 has an enriched chemistry with higher abundances of nitrogen-bearing molecules and carbon chain molecules (see \citealt{Sadavoy12}), where the former are generally associated with the freeze-out of \CO\ and its isotopologues, and the latter are generally thought to form on the surfaces of dust grains with carbon-rich icy mantles (e.g., \citealt{difran07}; \citealt{BerginTafalla07};  \citealt{Herbst09}), though recent studies have suggested that gas-phase productions of carbon-chain molecules are non-negligible (e.g., \citealt{Balucani15}).

Based on the combination of lower \COO\ abundances and more pronounced nitrogen-rich chemistry, B1-E2 appears to be uniquely depleted in \COO\ gas by a factor of two.   Depletion processes, such as molecular freeze-out, will produce thicker ice mantles on the dust grains toward B1-E2 compared to the other substructures.  Thicker ice mantles should increase the B1-E2 dust opacities (e.g., \citealt{Ossenkopf94}), which will consequently lower the measured column densities and bring B1-E2's \XCOO\ values into agreement with those of the other substructures.   Nevertheless, B1-E2 has a very moderate decrease in N(\COO) - only a factor of two - and it isn't clear how much the ice mantles (and thus, the opacity) will be affected from such a small level of depletion.  Moreover, to have higher opacities from thicker ice mantles, one must still assume that B1-E2 has higher levels of \COO\ freeze-out compared to the other substructures.  Therefore, we  believe B1-E2 is depleted in \COO\ gas by at most a factor of two due to molecular freeze-out.

The onset of \COO\ depletion toward a prestellar core precursor has not been previously identified.  This depletion (a factor of two) is much weaker than what is typically detected toward dense prestellar cores (e.g., \citealt{Caselli99}; \citealt{Bergin02}; \citealt{FordShirley11}), and corresponds better to the depletion seen in the lowest density objects (\citealt{Lippok13}) or in very young sources (e.g., \citealt{TafallaSantiago04}).   Since B1-E2 is slightly more centrally condensed than the other substructures (e.g., see Figure \ref{b1e_colDen} and \citealt{Sadavoy12}), this distinction in chemistry may be tied to its density.  Alternatively, B1-E2 is also dynamically distinct from the other substructures, suggesting that \emph{the evolution of turbulence and chemistry in dense cores appears to be contemporaneous}, such that only those objects with low internal turbulence also experience depletions of their carbon-bearing molecules and enhancements of their nitrogen-bearing and carbon-chain chemistry.  We discuss this chemodynamical picture in more detail in Section \ref{picture}.

\subsubsection{Comparisons with Simulations}\label{compSim}

Simulations of core formation have the difficult task of creating small-scale structures within larger-scale clouds under a variety of physical conditions (e.g., turbulence, magnetic fields, chemical processes).  The high density structures that form at early times in simulations generally have supersonic turbulent motions (e.g., \citealt{BallesterosParedes03}; \citealt{LiNakamura04}; \citealt{ClarkBonnell05}; \citealt{VazquezSemadeni05}; \citealt{BallesterosParedes07}).  Most of these early objects are transient and are destroyed within a crossing time unless they are able to build up sufficient mass (e.g., through mergers, ambipolar diffusion) to form long-lasting structures (\citealt{ClarkBonnell05}; \citealt{VazquezSemadeni05}).     Although the B1-E substructures have supersonic non-thermal line widths as expected in some simulations for high-density objects at early times,   they also appear pressure-confined (see Section \ref{sources}) rather than transient.  Thus, the B1-E substructures are not at the earliest stages when dense, transient objects first condense out of the ambient material.  Rather, the B1-E substructures are likely persistent objects, that will eventually form into dense (prestellar) cores.  

The high density structures that form out of these simulations can also interact with each other, causing them to evolve dynamically separate from their surrounding cloud.  \citet{Klessen00} found that mutual gravitational interactions between dense structures could homogenize their relative velocities compared to their initial velocity distribution, and that these interactions become more effective as the dense structures gain more mass.  Thus, it is possible that core formation only proceeds when the small-scale, high density structures decouple from the cloud.  Nevertheless, the expected timescale for these interactions is very long, $\sim 7 t_{ff}$.   Assuming a typical density of $10^3\ \vol$\ for the B1-E clump, the expected free fall time is $t_{ff} = \sqrt{3\pi/32G\rho} \sim 1$ Myr, and a timescale of $\sim 7 t_{ff}$ is older than all the YSOs in Perseus (e.g., \citealt{Bally08}).   If core formation proceeds only after the early precursors decouple from the ambient cloud via mutual interactions, this process must occur more efficiently.   Alternatively, \citet{BaileyBasu12} found that material with different length scales (e.g., clumps and cores) form separately over different timescales.  Thus, the B1-E substructures may have started decoupled from their parent clump, such that they formed with more similar relative motions than the random motions within the B1-E.  Whichever mechanism decouples high density structures from their parent cloud (e.g., mutual interactions or in-situ), it must occur early in the formation of these objects. 

Some simulations have suggested that high density structures inherit their kinematic properties from the material out of which they fragmented (e.g., \citealt{Klessen00}).  In particular for the B1-E substructures, if we assume that their non-thermal \ammonia\ (1,1) line widths are due to turbulent motions (e.g., not multiple subcomponents; see Section \ref{compObs}), then their dense gas dynamics, as traced by the \ammonia\ (1,1) data, have very similar turbulent velocity dispersions as the bulk \COO\ gas.  While the B1-E substructures appear decoupled from their parent clump, their internal kinematics still resemble that of B1-E, suggesting that the timescale associated with a core losing its internal turbulence may be much longer than the timescale associated with decoupling dense structures from the ambient cloud.  

The substructures also likely inherited chemical properties from the B1-E clump.  Most of the B1-E substructures are chemically unevolved compared to dense (prestellar) cores in that they show neither \CCO\ or \COO\ depletion nor enriched nitrogen or carbon-chain chemistry.   \citet{TafallaSantiago04} attributed low CO depletion to very young objects, such that processes like ambipolar diffusion have insufficient time to affect the internal chemistry.  Similarly, \citet{NakamuraLi05} attributed low CO depletion levels to formation mechanisms like rapid turbulent compression, where an object forms too quickly for its carbon-bearing molecules to sufficiently deplete.   Although, we find no indication of \COO\ depletion in most substructures, we do not see a rich carbon chemistry (e.g., CCS; \citealt{Sadavoy12}) as predicted by \citet{NakamuraLi05}.  Thus, the B1-E substructures appear to be very young.

On the other hand, B1-E2 is the only source to show signs of slight \COO\ depletion and moderate enriched nitrogen-chemistry.  Since B1-E2 is also the least turbulent substructure, chemical processes may also depend on the local turbulence.  Previous studies have connected turbulence and cloud chemistry on scales of the mixing length, though the spatial extent of turbulent mixing is still debated (e.g., \citealt{CantoRaga91}; \citealt{Xie95}; \citealt{Rawlings97}; \citealt{YateMillar03}; \citealt{Martinell06}).  Most of these studies, however, examined turbulent transport across clouds, whereas for B1-E, we need to determine whether or not turbulent mixing can replenish the carbon-chemistry over the scales of a cold, supersonic ``core.''  In addition to turbulent mixing, shocks can replenish carbon-bearing molecules from icy grain mantles back into the gas phase (e.g., \citealt{Draine83}; \citealt{WilliamsHartquist84}; \citealt{Guillet11}).  At the expected shock velocities of $2-3$ \kms\ (\citealt{Pon14}), dust grain mantles are unlikely to be entirely removed (e.g., \citealt{Draine83}).  Instead, gas-grain interactions involving heavier molecular gases like CO can sputter dust grains or their icy mantles at these lower velocities (\citealt{JimenezSerra08}).   In the sputtering case, a smaller fraction of gas-grain interactions will result in CO freeze-out.  Moreover, sputtering of dust grains with icy mantles may release additional CO molecules (e.g., those molecules that did freeze-out) back into the gas phase, thereby keeping a quasi-equilibrium gas phase abundance.  Thus, the supersonic B1-E substructures may have higher \COO\ abundances due to dust sputtering interactions, whereas the transonic B1-E2 source does not have the kinematics for sputtering, and as such, its dust-grain interactions primarily result in freeze-out.   Both chemical mixing and small-scale shock replenishment are possible explanations for the chemistry in the B1-E sources.  Detailed chemical models at both clump- and core-scales in a supersonic turbulent environment with low-velocity, small-scale shocks are needed to study the adsorption of CO ices and the transport of carbon-rich material to core scales.

\subsubsection{Connecting Dynamics and Chemistry}\label{picture}

Our observations suggest the following chemodynamical picture for the formation of relatively isolated prestellar cores; (1) core precursors condense out of their turbulent parent clump with their kinematic and chemical properties inherited from that medium.  These core precursors either form decoupled from their surroundings or are promptly decoupled, e.g., via efficient mutual gravitational interactions; (2) the highly turbulent core precursors dissipate their turbulence over some time, e.g., via small-scale shocks, while retaining their carbon-rich chemistry; (3) as the turbulence dissipates and the density increases, carbon-bearing molecules become depleted from the gas phase enriching nitrogen-bearing molecules and large carbon-chain molecules; (4) the core precursors continue to evolve on this path of turbulence dissipation, carbon depletion, and chemical enrichment, eventually forming a quiescent, chemically layered prestellar core.  Whether the turbulence dissipation and chemical depletion/enrichment occur simultaneously or after a threshold low turbulence level is reached is still unclear.  Thus, most of the B1-E substructures are in the above Stage 2, since they have not yet dissipated enough turbulence to allow for significant \COO\ depletion (i.e., they are still carbon-rich).  The B1-E2 substructure, however, is slightly more evolved in Stage 3.  The \COO\ depletion and chemical enrichment detected toward this object are both weak compared to most prestellar cores, indicating that B1-E2 is still very young.  Thus, the B1-E substructures represent excellent candidates for core precursors and a very good opportunity to catch the initial stages of core formation.

\section{Conclusions}\label{conc}

We have presented new \CCO\ and \COO\ observations in the $J=1-0$, $J=2-1$, and $J=3-2$ transitions of the Perseus B1-E clump.  To compare all data equally with the \emph{Herschel}-derived dust temperatures and column densities, we convolved the molecular line observations to a common resolution of 36.3\arcsec.  We compared and contrasted the kinematics of the entire B1-E clump with those of the individual nine substructures identified by \citet{Sadavoy12}.  Additionally, we characterized the \CCO\ and \COO\ chemical signatures.  Our main conclusions are:

\begin{enumerate}

\item{The B1-E clump has a velocity gradient of $\sim 1\ \kmspc$\ detected in both the \CCO\ (1-0) and \COO\ (1-0) spectra.  This velocity gradient indicates that the gas velocity is increasing from the NW to the SE at an angle of $\sim 20\degree-30$\degree, north to west.  In contrast, material in the larger Perseus cloud has velocities generally increasing from the SW to the NE.  Since differences in gas flows can lead to cloud interactions, the  formation of substructures may be triggered in B1-E by the shear of these different flows or by compression if the flows are colliding.  At this time, the origin of this different velocity gradient and its impact on B1-E is unclear.}

\item{We find that the B1-E substructures have more homogenized motions than typical turbulent velocities detected in the larger-scale ambient clump (as traced by \COO), similar to what has been found in other clumps in Perseus (e.g., see \citealt{Kirk09}; \citealt{Kirk10}).  Moreover, the \ammonia\ emission does \emph{not} show a velocity gradient.  Thus, the B1-E substructures appear decoupled from their environment, indicating that dense cores either form decoupled from their parent clump or obtain such homogenized motions through external factors early in their evolution.}

\item{We find that the turbulence in B1-E is difficult to quantify from the \CCO\ and \COO\ spectra as both may contain multiple velocity components.  Using single Gaussian fits, the median non-thermal velocity dispersion of the \COO\ (1-0) spectra is $\sim 0.4$ \kms, which agrees well with the broadest \ammonia\ (1,1) emission from \citet{Sadavoy12}.  We propose that core precursors form with the same turbulent properties as their parent clump.  As these sources evolve, their turbulence dissipates (e.g., via small-scale shocks; see \citealt{Pon14}) and the core precursors become quiescent.  Nevertheless, high resolution observations are necessary to determine whether the broad line emission toward the B1-E substructures are due to supersonic turbulence or multiple subcomponents along the line of sight.}  

\item{Using the radiation transfer code, RADEX, we find that B1-E is best fit with $\XCOO\ \sim 2 \times 10^{-7}$.   For $\NHH \lesssim 8 \times 10^{21}$ \cden, we see primarily lower values of \XCOO, suggesting that selective dissociation of \COO\ is significant at these column densities.  At $\NHH \gtrsim 10^{22}$ \cden, we find that the \COO\ column density toward the B1-E2 substructure is uniquely lower than the other sources.  We consider B1-E2 to be depleted in \COO\ by a factor of two, in agreement with its relatively enriched chemistry, as seen by \citet{Sadavoy12}.}

\item{Most B1-E substructures are undepleted in \COO\ and without enriched chemistry.  The exception, B1-E2, is also the only substructure with quiescent (transonic) turbulence.  Therefore, we suggest that the depletion of carbon-bearing molecules and the enrichment of nitrogen-bearing molecules are concurrent with turbulence dissipation, where cores themselves evolve from highly turbulent, undepleted objects to less turbulent objects with CO-depleted chemistry.}

\end{enumerate}

The B1-E clump is an excellent laboratory to explore core formation.  Since this region has not formed a previous generation of cores or protostars, we have the unique opportunity in B1-E to observe the initial conditions associated with dense cores condensing out of their parent clump.  In particular, our observations have hinted at a number of interesting possibilities, including possible triggering (from shearing or colliding flows) and an evolutionary sequence connecting turbulent dissipation to chemical enrichment in the core precursors.   Nevertheless, further study of this region is necessary to understand better the processes influencing this clump.  In particular, simulations that explore the core formation process and chemical depletion in a highly turbulent environment are key to connecting the initial characteristics of these objects to the characteristics of their surrounding clump environment.  Additionally, studies that explore the impact of low-velocity cloud interactions are necessary to determine the significance of a clump with a velocity flow different from that of its parent cloud.  With such studies, we may have, for the first time, the necessary tools to constrain core formation.

\vspace{1cm}
\acknowledgments{\noindent \emph{Acknowledgements:} This work was possible with funding from the Natural Sciences and Engineering Research Council of Canada PDF award.  We thank the anonymous referee for their comments that improved the clarity of this paper.  The authors thank N. Bailey, H. Beuther, P. Caselli, E. Keto, H. Kirk, J. Pineda, A. Pon, and D. Semenov for useful discussions.   This project utilized observations from \emph{Herschel}, the JCMT, the SMT, and the IRAM 30m telescope.   \emph{Herschel} is an ESA space observatory with science instruments provided by European-led Principal Investigator consortia and with important participation from NASA.     The JCMT is operated by the Joint Astronomy Centre (JAC) on behalf of the Science and Technology Facilities Council (STFC) of the United Kingdom, the National Research Council (NRC) of Canada, and the Netherlands Organisation for Scientific Research.      The SMT is operated by the Arizona Radio Observatory (ARO), Steward Observatory, University of Arizona, with support through the NSF University Radio Observatories program (URO: AST-1140030).     IRAM is supported by INSU/CNRS (France), MPG (Germany) and IGN (Spain).  }

\bibliographystyle{apj}
\bibliography{references}

\appendix

\renewcommand*\thetable{\Alph{section}.\arabic{table}}
\numberwithin{table}{section}
\renewcommand*\thefigure{\Alph{section}.\arabic{figure}}
\numberwithin{figure}{section}

\section{Justification of Dust Temperature}\label{TexAppendix}

In Section \ref{radex}, we used the radiative transfer code, \emph{RADEX}, to model the multi-J \CCO\ and \COO\ observations using the \emph{Herschel}-derived dust temperature as a proxy for the kinetic gas temperature.  Since the \CCO\ and \COO\ line emission mainly trace material at $n \lesssim 10^4$ \vol, the gas and dust may be decoupled such that the kinetic gas temperatures (at the densities traced by \CCO\ and \COO) are warmer than the dust temperatures (e.g., \citealt{Goldreich74}; \citealt{Goldsmith01}; \citealt{difran07}; \citealt{Ceccarelli07}).  Morevoer, the best-fit radiative transfer models had different excitation temperatures for each J-line, suggesting the molecules whose emission we observe are not in thermal equilibrium with the unknown kinetic gas temperature.  

We estimated the kinetic gas temperature using the COMPLETE survey (\citealt{Ridge06}), which mapped the entire Perseus molecular cloud in \CO\ (1-0) and \CCO\ (1-0) at $\sim 45$\arcsec\ resolution.  Using these data, \citet{Pineda08} determined the excitation temperature, \Tex, from the \CO\ (1-0) data assuming optically thick emission, following;
\begin{equation}
\Delta T_{MB} = (1-e^{-\tau})\frac{h\nu}{k}\left[f(\Tex) - f(T_{CMB})\right],
\end{equation}
where $\Delta T_{MB}$ is the main beam brightness of the line, $\tau$ is the optical depth, $h$ is the Planck constant, $k$, is the Boltzmann constant, and $f(T)$ corresponds to,
\begin{equation}
f(T) = \left[\exp\left(\frac{h\nu}{kT}\right)-1\right]^{-1}
\end{equation}
and $f(T)$ is measured for the excitation temperature (\Tex) and the background temperature, given by the cosmic microwave background (CMB).  Assuming a CMB temperature of 2.73 K, \Tex\ is measured as,
\begin{equation}
\Tex = \frac{5.53 \mbox{ K}}{\ln\left(1 + \frac{\mbox{5.53 K}}{\mbox{$\Delta T_{MB} + 0.84$ K}}\right)}.\label{eqTex}
\end{equation}

\citet{Pineda08} reported typical \Tex\ values of $\Tex \sim 11$ K in different regions of Perseus at 5\arcmin\ resolution (to match the resolution of their extinction data).  We followed the same procedure to characterize \Tex\ on a pixel-by-pixel basis at the native resolution of the COMPLETE data ($\sim$ 45\arcsec).  We used single Gaussian fits to measure the peak line brightness (\Tmb) at each pixel, assuming a beam correction factor of $\sim 0.45$ to convert from $T_A^{\star}$ to \Tmb.  We also assumed the emission filling fraction is unity.

\begin{figure}[h!]
\includegraphics[scale=0.49]{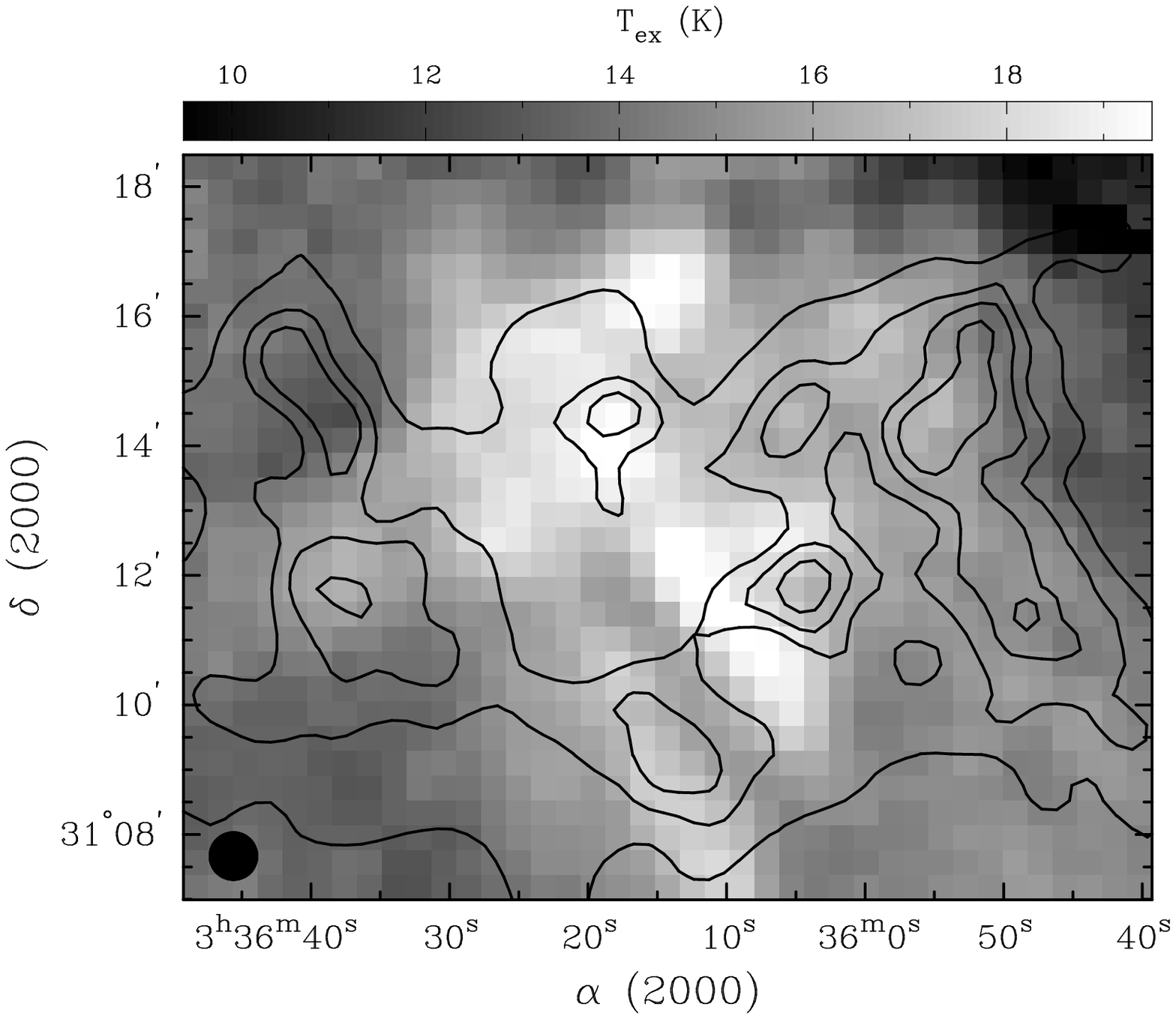} \includegraphics[scale=0.49]{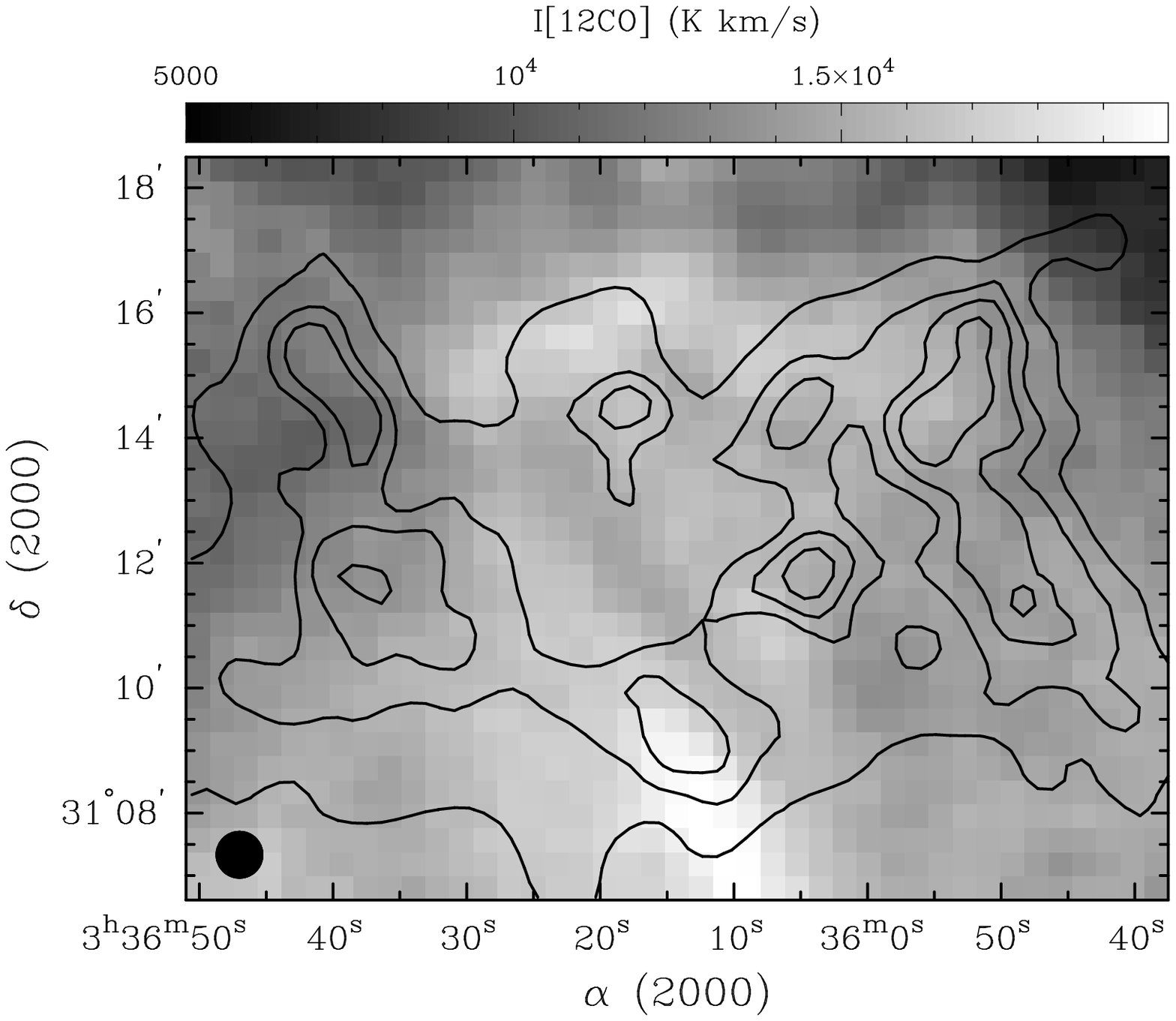} \\
\includegraphics[scale=0.49]{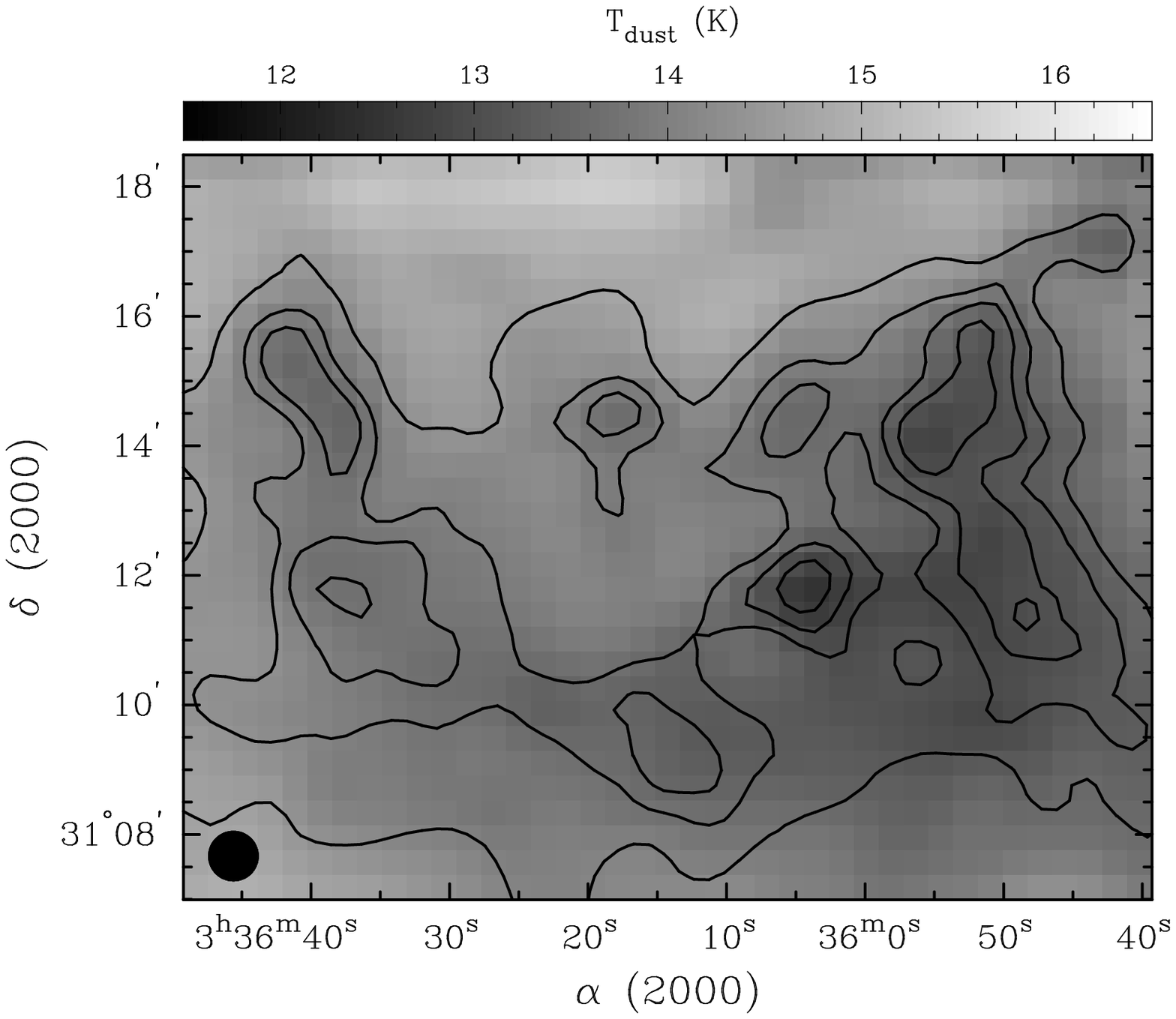} \includegraphics[scale=0.51]{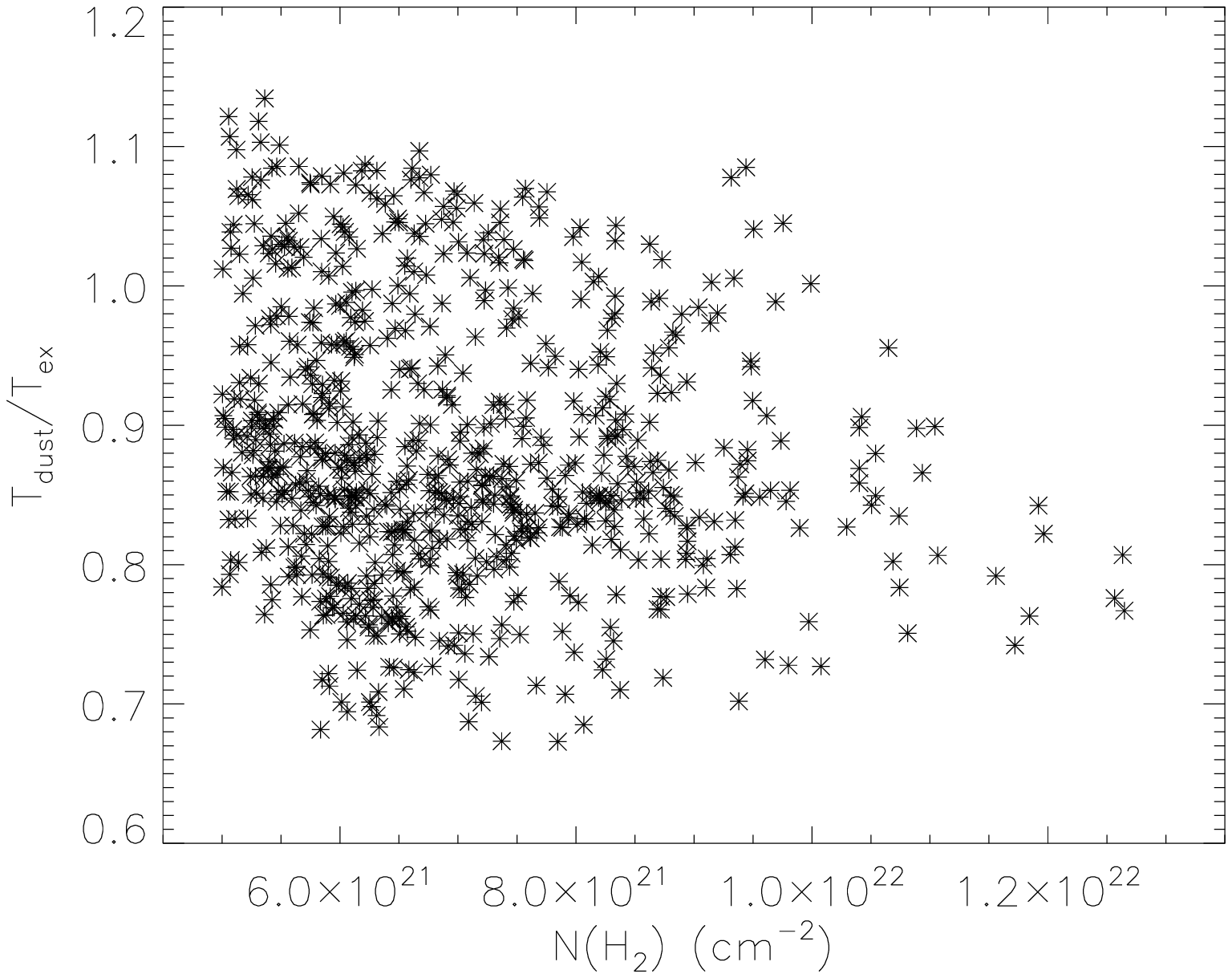}
\caption{\emph{Top:} Excitation temperature (\Tex) toward B1-E (left) and integrated intensity of \CO\ (1-0) line emission (right).  Both maps are at 46 \arcsec\ resolution and used data from the COMPLETE survey (\citealt{Ridge06}). \emph{Bottom:} \emph{Herschel}-derived dust temperature ($T_{dust}$) at 46\arcsec\ resolution (right) and the ratio of $T_{dust}$ and \Tex\ with column density (left).  We used modified blackbody SED fitting to determine $T_{dust}$ from $160-500$ \um\ \emph{Herschel} data assuming a dust emissivity index of $\beta = 2$.  The $T_{dust}$ map has also been convolved to 46\arcsec\ resolution to match the resolution of the \CO\ (1-0) data.  Contours show \emph{Herschel}-derived column density levels from HGBS data of  $5.0 \times 10^{21}\ \cden, 7.0 \times 10^{21}\ \cden, 8.5 \times 10^{21}\ \cden, \mbox{and}\ 10.5 \times 10^{21}$ \cden.}\label{tempComp}
\end{figure}

Figure \ref{tempComp} shows the excitation temperature across B1-E determined from Equation \ref{eqTex} with the integrated intensity map of the \CO\ (1-0) observations from COMPLETE.  For comparison, we also show the \emph{Herschel}-derived dust temperature ($T_{dust}$) map from SED fitting (see Section \ref{data} and \citealt{Sadavoy14} for details) and the ratio of $T_{dust}$ and \Tex.  Both temperature maps show very different structures.  The \CO-derived \Tex\ measurements range over $10-22$ K, with the highest values towards the center of the field, and no correlation is found between \Tex\ and column density (contours).  Since the \CO\ (1-0) line emission is optically thick, it will not trace the dust emission (e.g., see the \CO\ integrated intensity map in Figure \ref{tempComp}).  In contrast, the \emph{Herschel}-derived dust temperatures range over $12-16$ K, with a strong correlation between $T_{dust}$ and column density such that the lowest temperature values are towards the densest regions of B1-E.  Similarly, the \CCO\ and \COO\ integrated intensity maps also trace the dust emission well (see Figures \ref{integInt_13co} and \ref{integInt_c18o}).  Therefore, $T_{dust}$ is likely a better representation of the kinetic gas temperature associated with the \CCO\ and \COO\ gases than the \CO-derived \Tex\ values.  

Although the gas and dust are likely not coupled toward most of B1-E, the kinetic gas temperature and the dust temperature are expected to vary by only a few Kelvin at densities of $n \sim 10^{3-4}$ \vol\ (e.g., \citealt{Young04}; \citealt{Ceccarelli07}), which represent the bulk densities of B1-E (see \citealt{Sadavoy12}).  Therefore, we re-ran the RADEX models for a randomly selected sample of test pixels assuming the CO gas kinetic temperature is warmer than the dust temperature by 2 K.  We found that the \CCO\ column densities were generally lower by $\lesssim$ 20\%, whereas the \COO\ column densities were lower by $< 10$\%.  These uncertainties are relatively minor, considering the complex spectra of the \CCO\ and \COO\ gases and the simple geometric assumptions in the RADEX routine.  As such, in the absence of proper gas kinetic temperatures, the \emph{Herschel}-derived dust temperatures give a good first-order estimate for abundances from RADEX.

\end{document}